\documentclass{aa-package/aa}

\usepackage{graphicx}

\begin{document}

\title{Physical parameters of 15 intermediate-age LMC clusters from modelling 
of HST colour-magnitude diagrams}

\author{L. O. Kerber\inst{1} \and B. X. Santiago\inst{2} \and E. Brocato\inst{3}}

\offprints{Leandro Kerber, \email{kerber@astro.iag.usp.br}}

\institute{Universidade de S\~ao Paulo, Rua do Mat\~ao 1226, 
Cidade Universit\'aria, S\~ao Paulo, 05508-900, SP, Brazil
\and Universidade Federal do Rio Grande do Sul, IF, CP 15051, Porto Alegre,
91501-970, RS, Brazil \and
INAF - Osservatorio Astronomico di Collurania, Via M. Maggini, 64100 Teramo, Italy}

\date{Received 27 July 2006 / Accepted 3 October 2006}

\abstract
% context heading (optional){}
{}
% aims heading (mandatory)
{We analyzed HST/WFPC2 colour-magnitude diagrams (CMDs) of 15 populous Large
Magellanic Cloud (LMC) stellar clusters with ages between $\sim$ 0.3 Gyr 
and $\sim$ 3 Gyr.  
These (V, B-V) CMDs are photometrically homogeneous and typically 
reach V$\sim$22.
Accurate and self-consistent physical parameters 
(age, metallicity, distance modulus and reddening) were extracted 
for each cluster by comparing the observed CMDs with synthetic ones.}
% methods heading (mandatory)
{These determinations involved simultaneous statistical 
comparisons of the main-sequence fiducial line and the red clump position, 
offering objective and robust criteria to determine the best models. 
The models explored a regular grid in the parameter space covered
by previous results found in the literature. Control experiments
were used to test our approach and to quantify formal uncertainties.}
% results heading (mandatory)
{In general, the best models show a satisfactory fit 
to the data, constraining well the physical parameters of each 
cluster. 
The age-metallicity relation derived by us presents a lower 
spread than similar results found in the literature for the same clusters.
Our results are in accordance with the published ages
for the oldest clusters, but reveal a possible underestimation
of ages by previous authors for the youngest clusters. 
Our metallicity results in general agree with the ones
based on spectroscopy of giant stars and with recent works 
involving CMD analyses.   
The derived distance moduli implied by the most reliable solutions, 
correlate with the reddening values, as expected from the non-negligible
three-dimensional distribution of the clusters within the LMC.}
% conclusions headind (optional)
{The inferred spatial distribution for these clusters is roughly aligned 
with the LMC disk, being also more scattered than recent numerical 
predictions, indicating that they were not formed in the LMC disk.
The set of ages and metallicities homogeneously 
derived here can be used to calibrate integrated light studies
applied to distant galaxies.
}


\titlerunning{Physical parameters of 15 intermediate-age LMC clusters}
\authorrunning{Kerber et al.}

\maketitle

\section{Introduction}
\label{Intro}

The Large Magellanic Cloud (LMC) is a useful ensemble of stars and stellar 
systems, since it is a galaxy with remarkably distinct characteristics
when compared with to Galaxy, while its distance is close enough so that its
stellar content is well resolved (Olszewski et al. 1996; Westerlund 1997). 
This rich information imprinted in the LMC includes 
its large system of more than 1800 identified stellar clusters 
(Bica et al. 1999). Some rich LMC clusters may be as old as Milky Way globular
clusters; many others have ages similar to those inferred for the 
open clusters in the disk of our Galaxy, but are generally richer and more
metal-poor than these are.
Therefore, studies of individual LMC clusters, as well as of its
entire cluster system, have lead to valuable contributions to the
understanding of how clusters, and the stars within them, form and
evolve.

Many examples can be found in the recent literature that illustrate
this promising field. With respect to the impact on stellar evolution
theory, evolutionary tracks and isochrones of young and subsolar 
metallicity stars are being continuously tested, 
resulting in stimulating discussion about the efficiency of the convective overshooting process
(Brocato et al. 2003; Gallart et al. 2003).
The spatial variation of the stellar luminosity and mass functions 
observed in clusters has helped us better understand the mass 
segregation effect 
(de Grijs et al. 2002; Gouliermis et al. 2004; Kerber \& Santiago 2006)
and has thus contributed to the discussion about the IMF universality.
In terms of the cluster systems, the lack of populous LMC clusters with ages 
between 4 and 10 Gyr (the so-called ``age gap''; the only known
exception is ESO 121-SC03), 
also imprinted on the cluster age-metallicity relation (AMR) 
(Olszewski et al. 1991; Bica et al. 1998; Geisler et al. 1998), 
was recently reproduced by numerical N-body simulations
with realistic conditions for the clusters formation    
that take into account the interaction between LMC, the Small
Magellanic Cloud (SMC) and the Galaxy 
(Bekki et al. 2004; Bekki \& Chiba 2005).

LMC clusters also provide a decisive contribution to the
calibration of models describing the integrated light (spectra and colours) 
of single stellar populations (SSP). These models in turn are crucial
for the studies of distant unresolved stellar populations.
Therefore, it is necessary that these models 
recover ages and metallicities of LMC clusters which are in agreement 
with those obtained by methods that rely on the analysis of resolved
stars. Otherwise there is a serious 
risk that the interpretation of the stellar content of unresolved galaxies 
will be severely biased. Several recent works either fully 
or partially study the integrated light of LMC clusters:
Leonardi \& Rose (2003), Santos \& Piatti (2004), Santos et al. (2006),
de Grijs \& Anders (2006), McLaughlin \& van der Marel (2005),
Beasley et al. (2002), Goudfrooij et al. (2006).
The results of all these studies on the LMC clusters are 
founded on two observational pillars: the spectroscopy of individual 
red giants and the photometry of dense systems that make up
colour-magnitude diagrams (CMDs). 
While the former provides metallicities, largely based on the 
calcium triplet lines, the latter yield ages, 
metallicities, distance modulus and reddening values. 

Spectroscopic studies include those by Olszewski 
et al. (1991) (hereafter OSSH), which determined [Fe/H] for 
$\sim$ 70 LMC clusters, and Cole et al. (2005), which did
the same for 373 LMC field stars.   
Recently, Geisler (2006) (see also Grocholski et al. 2006) 
showed new results for 29 clusters 
observed with the FORS2 instrument on the 
Very Large Telescope (VLT), where typically 8 red giants
per clusters were used to constrain the metallicity of each
cluster. 
This latter work, that exceeds the OSSH in quality but 
covers a lower number of clusters, is a good example of the
successful application of multi-object spectroscopy (MOS) 
for LMC clusters.

As for CMD analysis, it has been used as a powerful tool to
determine the physical parameters of stellar systems as well as
to calibrate stellar evolution theory.
CMDs based on {\it Hubble Space Telescope} (HST) or on 8m class 
telescopes, coupled with detailed analysis techniques, have allowed 
accurate determinations of age and metallicities for 
LMC clusters (Kerber \& Santiago 2005; Bertelli et al. 2003, Woo et al. 2003)
and star formation history (SFH) for neighbouring galaxies,
including the LMC (Gallart et al. 1999; Dolphin 2002; Javiel et al. 2005).
A common feature of the aforementioned works is that they combine synthetic 
CMDs (generated by numerical simulations) with statistical 
tools to discriminate the best models, constituting
a testable and objective approach to 
recover the physical information from an observed CMD.
This kind of study, if applied to a large number of clusters,
should significantly improve the age determinations based 
on lower-resolution data (e.g., Elson \& Fall 1988; Girardi et al. 1995).
For instance, Dirsch et al. (2000), Geisler et al. (2003), 
Piatti et al.(2003ab) are examples of recent works that applied
homogeneous analyses to CMDs taken from a sizable number of LMC 
clusters, although their data still come from ground-based 
small telescopes.

With this in mind we analyzed a sample of HST data taken with
the {\it Wide Field and Planetary Camera 2} (WFPC2)
CMDs of populous LMC clusters published by Brocato et al. (2001). 
We selected the 15 intermediate-age clusters (IACs) in this sample
that satisfy the following criteria: i) inferred age 
between $\sim 0.3$ to $\sim 3.5$ Gyr; ii) CMDs that display a 
main-sequence (MS) stretching at least 1 magnitude below the turn-off
(MSTO) point and with evidence of red clump (RC) stars. 

These data are photometrically homogeneous and
typically reach V $\sim 22$ for stars in the cluster´s centre.
The main goal of this work is to provide ages, metallicities,
distance moduli and reddening values for each cluster 
in a self-consistent method based on a homogeneous
and robust analysis. 
Therefore, these physical parameters can be very useful 
for the calibration of integrated light SSP models.
Furthermore, the set of derived distance moduli for individual 
clusters offers a good opportunity to probe the three-dimensional
distribution of the intermediate-age clusters within the LMC,
adding new information and important constraints to the understanding
of the stellar cluster formation in this neighbouring galaxy.

In the next section we present the observed CMDs and the
cluster sample.
The CMD modelling process is presented in Sect. \ref{modelling}
while the model grid and the previous determinations found in the 
literature are shown in Sect \ref{grid}.
Sect. \ref{stattools} is dedicated to the statistical tools that 
objectively discriminate among the best models.
The results are presented and discussed in Sect. \ref{results}; 
we first discuss the results on a cluster-by-cluster basis, but also
investigate the sample properties as a whole.
The last section shows the conclusions and the summary.

\section{The data}
\label{data}

\begin{figure*}
\centering
\includegraphics[width=17cm]{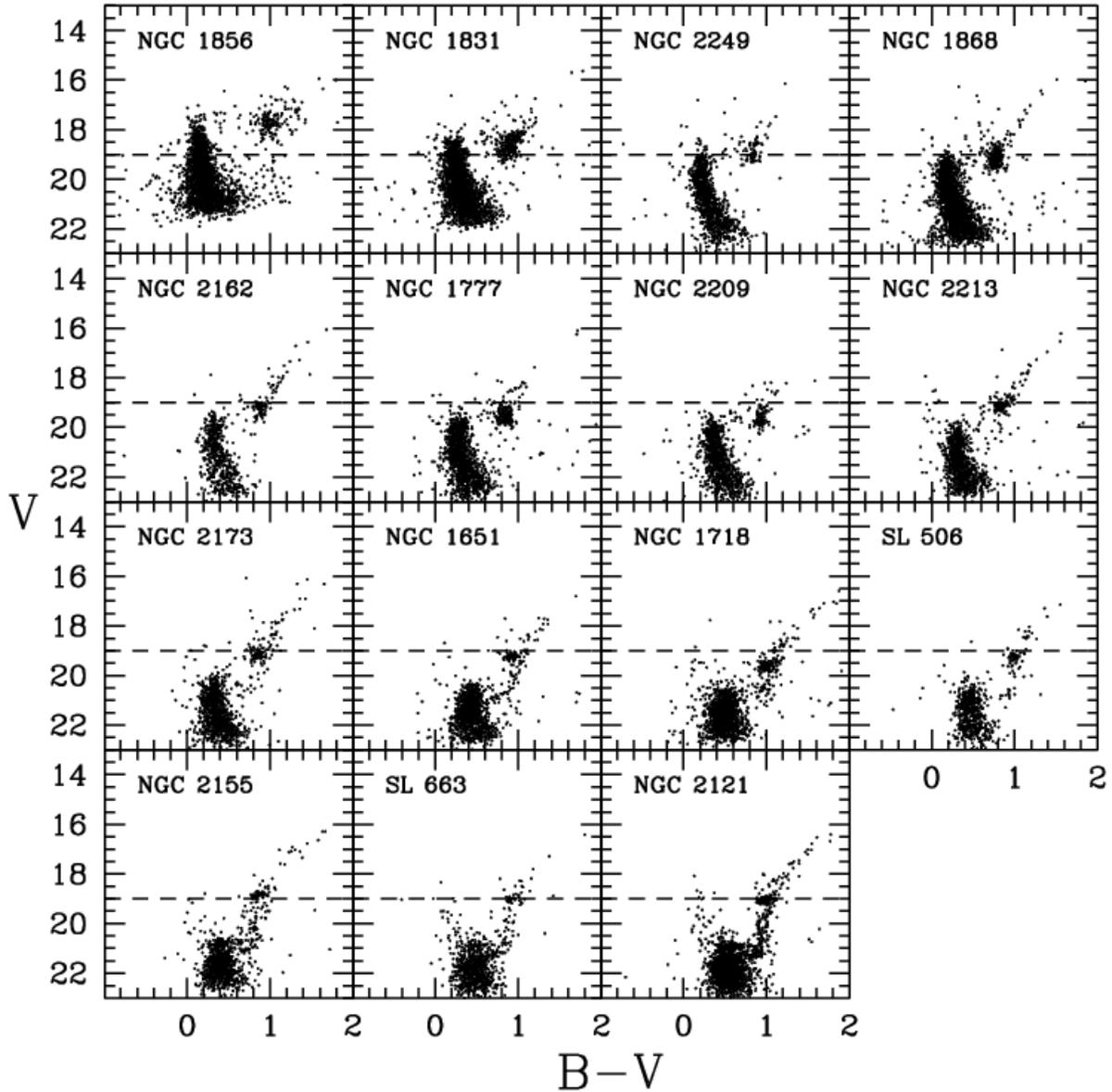}
\caption{CMDs for 15 intermediate-age LMC clusters by Brocato et al
(2001); only stars with R $\leq 2~R_{c}$ (R $\leq 1~R_{c}$ for NGC
  2121) are shown.
This sequence was suggested by those authors to reveal the ``evolution'' 
effect in the sample. The horizontal dashed line at V=19.0 corresponds 
to the MS end for a cluster with $\tau \sim 1.0$ Gyr at the LMC
distance; it also roughly corresponds to the V magnitude of RC stars in the oldest clusters. }
\label{allcmds}
\end{figure*}

The data used in the present work were taken with the HST/WFPC2
for the following 15 rich intermediate-age LMC stellar clusters:
\object{NGC\,1651}, \object{NGC\,1718}, \object{NGC\,1777}, 
\object{NGC\,1831}, \object{NGC\,1856}, \object{NGC\,1868}, 
\object{NGC\,2121}, \object{NGC\,2155}, \object{NGC\,2162},
\object{NGC\,2173}, \object{NGC\,2209}, \object{NGC\,2249}, 
\object{SL\,506} and \object{SL\,663}.
These data were selected from the HST archive and reduced 
by Brocato et al. (2001). Their original 
cluster sample has 21 LMC clusters and one SMC cluster, covering a wider 
range in age ($0.1 \la \tau \la 13$ Gyr).
For each cluster, snapshot images were obtained in the F450W 
($\sim B$) and F555W ($\sim V$) filters, and 
photometrically transformed to the standard system.
The HST archival images, data reduction process, procedures to calibrate 
the data and to evaluate the incompleteness are all detailed in
Brocato et al. (2001). 

The final (V, B-V) CMDs from Brocato et al. (2001) for the 15 
intermediate-age LMC clusters are presented in Fig. \ref{allcmds}.
These CMDs include only central stars, therefore 
reducing significantly the contamination by field LMC stars. 
We adopt a cut-off radius of $R \leq 2~R_{c}$, where $R_c$ is the
core radius for all clusters, except NGC\,2121. 
This latter is located in one of the most contaminated 
directions towards the LMC and has one of the faintest MS
terminations;  we thus selected inner stars ($R \leq 1~R_{c}$) 
to avoid field star contamination.
These are the CMDs we used to extract the 
physical parameters of each cluster (see Sect. \ref{results}).

We notice two common features in the CMDs: 
the extended MS ($V\sim 22$) and the prominent presence of 
stars in the helium-burning RC phase. 
In this figure they follow the same sequence as suggested by 
Brocato et al. (2001) to reveal the age effect in a LMC cluster.
As the cluster becomes older, both the MS termination and the RC
become less bright. However, this latter stalls at V $\sim$19.0.
Thus the $V_{MSTO} - V_{RC}$ magnitude difference increases
for clusters older than $\sim 1$ Gyr, as a consequence being a good age 
indicator (Geisler et al. 1997; Castellani et al. 2003). 
Additional features that can be seen in the CMDs are the red giant
branch (RGB) and, for the older clusters, the sub-giant branch (SGB).

\begin{figure}
\resizebox{\hsize}{!}{\includegraphics{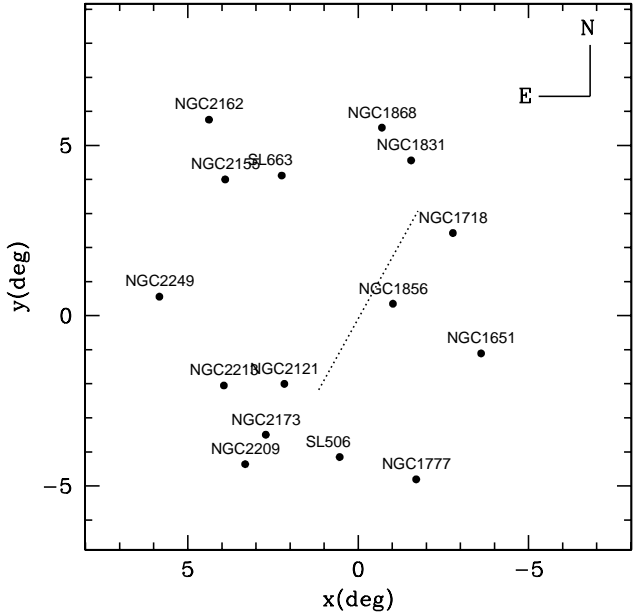}}
\caption{ 
Distribution on the sky of the LMC clusters (solid circles) 
analyzed in this work.
The dotted line indicates the line of nodes of the LMC disk, 
as determined by Nikolaev et al. (2004), while the 30 Dor position
is shown by an open circle.
The positions are relative to the optical centre of the LMC bar,
at RA$=05\fh 20\fm 56\fs$, DEC$=-69\degr 28\arcmin 41\arcsec$ (J2000.0)
(Bica et al. 1996).
}
\label{LMCsky}
\end{figure}

The on-sky distribution of our cluster sample is shown in Fig. \ref{LMCsky},
together with the 30 Dor position and the line of 
nodes of the LMC disk, as determined by Nikolaev et al. (2004).
The clusters have distances to the optical 
centre of LMC bar typically between $\sim 3\degr$ and $\sim 6\degr$, 
being spread in every quadrant with respect to this centre, but preferentially
located in the east side of the LMC.

\section{CMD modelling}
\label{cmdmodel}

To model a CMD we follow the numerical approach similar to that 
used by several authors in previous studies of resolved stellar 
populations (Gallart et al. 1999; Dolphin 2002; Bertelli et al. 2003; 
Woo et al. et al. 2003; Kerber \& Santiago 2005).
We generate synthetic CMDs that reproduce as accurately as possible 
the main features found in the observed CMD,
finding the best model CMDs based on statistical comparisons in the CMD plane.
The method is applicable to composed stellar populations (CSP), such as 
field stars in the Galaxy or in neighbouring galaxies, or to SSPs,
as in the present case.

We thus modelled the CMDs of the LMC clusters, considering that each of them 
is an SSP, characterized by stars with the same age ($\tau$) and 
metallicity ($Z$) (described by a Padova isochrone, Girardi et al. 2002), 
placed at the same distance ($(m-M)_{0}$) and subjected to the same
reddening ($E(B-V)$). 
These parameters uniquely define the position in the
(V,B-V) plane for single stars of a given mass.
The number of stars in a given CMD position is fixed by the 
stellar mass distribution, more commonly referred to as the
Present Day Mass Function (PDMF). It is parameterized here by a 
power law ($dN/dm \sim m^{-\alpha}$).
However, real observations suffer from photometric uncertainties and 
the effect of unresolved binaries (or blending of stars). 
The former effect is modelled using the photometric errors measured 
from the data. The effect of unresolved binaries is introduced by combining the
fluxes of two synthetic stars in a fraction ($f_{\rm{bin}}$) of the CMD points.
Only binaries whose secondary mass ($m_{2}$) is at least 70\% of the 
primary star mass ($m_{1}$) contribute to ($f_{\rm{bin}}$), the masses
being randomly selected according to a uniform mass ratio 
($q=m_{2}/m_{1}$) distribution.
Therefore, $f_{\rm{bin}}$ defined in this way incorporates only binaries
whose CMD position is different from that of the primary stars alone.

\begin{figure}
\resizebox{\hsize}{!}{\includegraphics{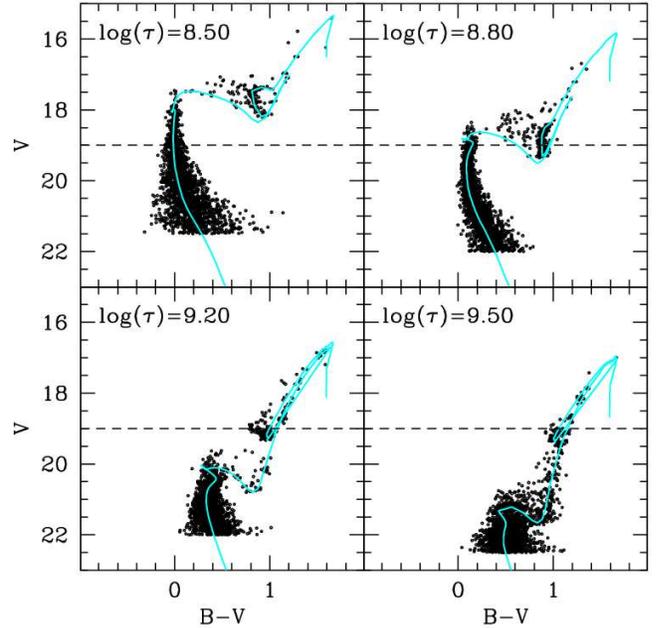}}
\caption{Examples of synthetic CMDs of rich LMC clusters 
following a sequence in age, as indicated in each panel.  
The other physical parameters for these CMDs were fixed: $Z=0.008$, 
$(m-M)_{0}=18.50$, $E(B-V)=0.05$, $\alpha=2.00$ and $f_{\rm{bin}}=30\%$.
The photometric uncertainties used in the models are consistent with
the ones observed in the Brocato et al. (2001) sample.
The horizontal dashed line at V=19.0 is a guideline for the MSTO
and the RC position, as explained in Fig. \ref{allcmds}.
}
\label{modelling}
\end{figure}

To illustrate our CMD modelling process, in Fig. \ref{modelling} we present
examples of synthetic CMDs of rich LMC clusters. 
The CMDs are disposed in an age sequence, approximately covering
the age range expected for the IACs from Brocato et al. (2001). 
As in Fig. \ref{allcmds}, the effect of age in the CMDs is noticeable.
The Padova isochrones used to generate the synthetic CMDs are also 
plotted in this figure. 
We deliberately adopt this set of stellar evolutionary models in this 
work because it presents the advantage of a thinner 
grid in age and metallicity than the others (see Sect. \ref{grid}), 
being also widely used and tested, in general offering good fits to the 
data.
Also the Padova isochrones adopt reasonable assumptions for convective 
overshooting, although Bertelli et al. (2003) found some evidence 
for a greater efficiency.

\section{Model grid and literature results}
\label{grid}

\begin{table*}
\caption{
Ages and metallicities summarized by Mackey \& Gilmore (2003) 
(first row for each cluster) 
and found in the literature for all clusters in our sample.
}
\label{lit_age_FeH}
\renewcommand{\tabcolsep}{2.0mm}
\centering
\begin{tabular}{l c c c c c c c l c c c c c c}
\hline\hline
Cluster & log($\tau$/yr) & ref & [Fe/H] & ref & ~~~~ & Cluster & log($\tau$/yr) & ref & [Fe/H] & ref\\ 
\hline
NGC\,1651 & $9.30^{+0.08}_{-0.10}$ &  5 & $-0.37\pm0.20$ & 12 & ~ & NGC\,2162 & $9.11^{+0.12}_{-0.16}$ &  5 & $-0.23\pm0.20$ & 12 \\
~	  & $9.34\pm0.08$ & 8 & $-0.82\pm0.44$ & 8 & ~ & ~  & $9.32\pm0.06$ & 8 & $-0.90\pm0.03$ & 8 \\
~	  & $9.26\pm0.08$ & 15 & $-0.07\pm0.10$ & 15 & ~ & ~ & $8.95\pm0.10$ & 6  &  & \\
~	  & $9.40$ to $9.10$ & 3 & $-0.63$ to $-0.45$ & 3 & ~ & NGC\,2173 & $9.33^{+0.07}_{-0.09}$ &  5 & $-0.24\pm0.20$ & 12 \\
NGC\,1718 & $9.30\pm0.30$ &  4 & $\sim -0.42$ & 10 & ~ & ~ & $9.19\pm0.04$ & 2 & $-1.05$ to $-0.80$ & 2 \\
~	  & $9.30^{+0.16}_{-0.14}$ & 1 & $-0.98^{+0.29}_{-0.30}$ & 1 & ~ & ~ & $9.62\pm0.05$ & 8 & $-1.38\pm0.08$ & 8 \\
~	  & $9.69^{+0.05}_{-0.07}$ & 1 & $-1.12^{+0.18}_{-0.22}$ & 1 & ~ & ~ & $9.18\pm0.08$ & 16 & $\sim -0.7$ & 16  \\
NGC\,1777 & $9.08^{+0.12}_{-0.18}$ &  5 & $-0.35\pm0.20$ & 12 & ~ & ~ & $9.06\pm0.10$ & 6  &  & \\
~	  & $9.01\pm0.05$ & 8 & $-0.39\pm0.01$ & 8 & ~ & NGC\,2209 & $8.98^{+0.15}_{-0.24}$ &  5 & $\sim -0.47$ & 10 \\
NGC\,1831 & $8.50\pm0.30$ &  4 & $+0.01\pm0.20$ & 12 & ~ & ~ & $\sim9.18$ & 11 &  &  \\
~	  & $8.70\pm0.03$ & 9 & $-0.20\pm0.10$ & 9 & ~ & ~ & $8.96\pm0.10$ & 6  &  & \\
~	  & $8.70\pm0.14$ & 8 & $-0.65\pm0.02$ & 8 & ~ & NGC\,2213 & $9.20^{+0.10}_{-0.12}$ &  5 & $-0.01\pm0.20$ & 12 \\
~	  & $8.60\pm0.10$ & 6  &  &  & ~ & ~ & $9.32\pm0.02$ & 8 & $-0.88\pm0.06$ & 8  \\
NGC\,1856 & $8.12\pm0.30$ &  4 & $\sim -0.52$ & 10 & ~ & ~ & $9.01\pm0.14$ & 7  &  & \\	 
~	  & $8.53^{+0.03}_{-0.13}$ & 1 & $-0.09^{+0.19}_{-0.10}$ & 1 & ~ & NGC\,2249 & $8.82\pm0.30$ &  4 & $\sim -0.47$ & 10 \\ 
~	  & $8.78^{+0.04}_{-0.08}$ & 1 & $-0.25^{+0.19}_{-0.18}$ & 1  \\
~	  & $8.50\pm0.14$ & 7  &  & & ~ & ~ & $8.44\pm0.30$ & 8 & $-0.40\pm0.02$ & 8  \\
NGC\,1868 & $8.74\pm0.30$ &  4 & $-0.50\pm0.20$ & 12 & ~ & ~ & $8.54\pm0.10$ & 6  &  & \\
~	  & $8.95\pm0.03$ & 9 & $-0.40\pm0.10$ & 9 & ~ & SL\,506   & $9.26^{+0.09}_{-0.11}$ &  5 & $-0.66\pm0.20$ & 12 \\
~	  & $8.97\pm0.04$ & 8 & $-0.32\pm0.71$ & 8 & ~ & ~ & $9.23\pm0.10$ & 9 & $-0.40\pm0.20$ & 9 \\
~	  & $8.87\pm0.10$ & 6  &  & & ~ & SL\,663   & $9.51^{+0.06}_{-0.07}$ &  13 & $-0.60\pm0.20$ & 12 \\
NGC\,2121 & $9.51^{+0.06}_{-0.07}$ & 13 & $-0.61\pm0.20$ & 12 & ~ & ~ &  &  & $-0.60\pm0.20$ & 13 \\
~	  &  &  &$-0.60\pm0.20$ & 13 & ~ & ~ & $\sim 9.60$ & 14 & $\sim -1.0$ & 14 \\
~	  & $9.40^{+0.08}_{-0.09}$ & 11 & $-0.65\pm0.20$ & 11 \\
~	  & $9.38\pm0.07$ & 11 & $-0.5\pm0.2$ & 11 \\
~	  & $9.60\pm0.14$ & 7  &  & \\
~	  & $\sim 9.60$ & 14 & $\sim -1.0$ & 14 \\ 
NGC\,2155 & $9.51^{+0.06}_{-0.07}$ &  13 & $-0.55\pm0.20$ & 12 \\
~	  &  &  &$-0.60\pm0.20$ & 13 \\
~	  & $\sim9.45$ & 2 & $-0.98$ to $-0.80$ & 2  \\
~	  & $9.43\pm0.26$ & 8 & $-0.44\pm0.86$ & 8  \\
~	  & $\sim9.56$ & 11 & $\sim-0.80$ & 11  \\
~	  & $9.46\pm0.05$ & 16 & $\sim-0.7$ & 16 \\
~	  & $9.45\pm0.14$ & 7  &  & \\
~	  & $\sim 9.60$ & 14 & $\sim -1.0$ & 14 \\ 
\hline
\end{tabular}

\noindent
Reference list (technique): 
1 - Beasley et al. (2002) (integrated spectra); 
2 - Bertelli et al. (2003) (VLT/CMD);
3 - Dirsch et al. (2000) (CMD); 
4 - Elson \& Fall (1988) (CMD); 
5 - Geisler et al. (1997) (CMD); 
6 - Girardi et al. (1995) (CMD);
7 - Girardi \& Bertelli (1998)(integrated colours); 
8 - Leonardi \& Rose (2003) (integrated spectra)
9 - Kerber \& Santiago (2005) (HST/CMD);
10 - Mackey \& Gilmore (2003) (crude estimation based the [Fe/H] from others clusters with similar ages);
11 - Piatti et al. (2003b) (CMD);
12 - Olszewski et al. (1991) (spectroscopy of red giants);
13 - Rich et al. (2001) (HST/CMD);
14 - Sarajedini (1998) (HST/CMD);
15 - Sarajedini et al. (2002) (CMD);
16 - Woo et al. (2003) (VLT/CMD)
\end{table*}

For each cluster, we explored a grid of models covering the parameter
space within reasonable limits. The ranges in age and metallicity in
the grid are consistent with the values compiled from the literature by 
Mackey \& Gilmore (2003) (hereafter MG03). 
These ranges are shown in the first row for each 
cluster in Table \ref{lit_age_FeH}; they are meant to homogeneously
cover the region in parameter space where the best models are expected
to lie and therefore allow accurate age and metallicity determinations.
All clusters have age results that come from CMD analyses,
although carried out by different authors and with variable data quality.
Among these, we quote Rich et al. (2001) based on HST data, 
and Geisler et al. (1997) who analysed homogeneous CMDs of 7 of 
our clusters. 
One of the pioneering works in LMC cluster age determination, 
Elson \& Fall (1988), completes the age list from MG03. 
The OSSH metallicity values, based on spectroscopy of
red giants, were assumed. 
For those clusters not observed by OSSH, MG03 estimated a crude
[Fe/H] value based on the metallicities of clusters with 
similar ages, and therefore these estimates are only 
guidelines and should be used with caution.    
The ages compiled by MG03 are not necessarily consistent with the 
metallicities determined by OSSH, 
as the former results come from CMD analyses that also assumed 
an independent, and often different, metallicity value.

Since the publication of the MG03 compilation, some new ages and metallicities
based on different techniques have been published.
These results are also listed in Table \ref{lit_age_FeH}, 
together with the ones from older works 
(Girardi et al. 1995; Girardi \& Bertelli 1998; Dirsch et al. 2000), 
which are still widely used. 
Also, we included the HST/CMDs analyses done by Sarajedini (1998) 
for the three oldest clusters in our sample, although he found 
solutions with overestimated ages and underestimated metallicities,
as discussed and demonstrated by Rich et al. (2003).
Recent works related to CMD analyses include 
(with the number of clusters in common with us and the 
telescope used shown in parenthesis):
Bertelli et al. (2003) and Woo et al. (2003) (2, VLT);
Piatti et al. (2003b) (3, CTIO 0.9m);
Kerber \& Santiago (2005) (3, HST/WFPC2).
Although less reliable than the CMD results, we also presented
the ones obtained from the analyses of integrated spectra done by 
Beasley et al. (2002) and Leonardi \& Rose (2003)
because they offer a good opportunity to check the 
consistency of the derived age and metallicity that comes from 
this kind of technique. 

\begin{figure}
\resizebox{\hsize}{!}{\includegraphics{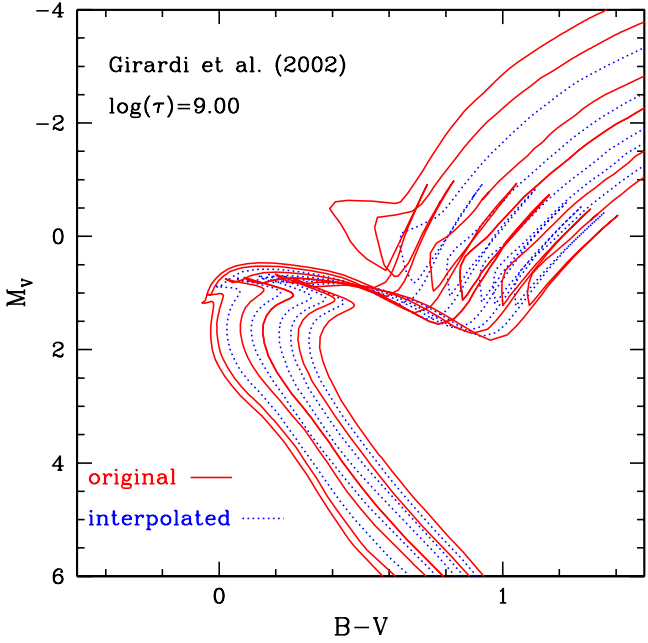}}
\caption{Original isochrones (solid lines) from Girardi et al. (2002)
and interpolated ones (dotted lines) using TRILEGAL code (Girardi et al. 2005).}
\label{isot}
\end{figure}

To reduce the discreteness in the model grid we adopted the smallest age step
published by Girardi et al. (2002),  $\Delta \rm{log}(\tau/yr)=0.05$,
and a thinner grid in Z than the original one, kindly provided 
by L. Girardi using the TRILEGAL code (Girardi et al. 2005). 
This metallicity grid, illustrated in Fig. \ref{isot}, is made up with 
Z=0.0001, 0.0004, 0.002, 0.004, 0.006, 0.008, 0.012, 0.016, 0.019 
($Z_{\odot}$), 0.024 and 0.030, where the new isochrones were obtained by 
interpolating between the original ones. 
To convert the Z values to [Fe/H], we assumed that 
[Fe/H]=$\rm{log}(Z/Z_{\odot})$.

Since reddening and distance modulus were considered free parameters
in the modelling, we explored ranges in the model grid that 
are compatible with the ones found for the LMC. 
The models span the range from $(m-M)_{0} = 18.20$ ($\sim 43.7$ kpc) to
$(m-M)_{0}=18.80$ ($\sim$ 57.5 kpc) (with a step of 0.05)
and from $E(B-V)=0.00$ to $E(B-V)=0.25$ (with a step of 0.01).
While the first range is consistent with a spherical distribution of clusters
with a radius of $\sim$ 9 kpc (roughly 10 deg on the sky) and centred at
the canonical LMC distance ($(m-M)_{0}=18.50$),
the second range is in accordance with the new reddening maps 
published for the LMC 
(Nikolaev et al. 2004; Zaritsky et al. 2004; Subramaniam 2005).
Although in some cases these authors derived E(B-V) values 
for directions close to the clusters, we preferred not 
to fix reddening for any cluster, since it can be located in the
foreground or background relative to the bulk of the stars considered in 
these works. 

The PDMF slope was fixed at $\alpha=2.00$, 
in agreement with the recent determinations for the central regions
of LMC clusters (Kerber \& Santiago 2006).
We adopted a typical value for the fraction of binaries
of $f_{\rm{bin}}=30\%$, in accordance with the determination done by 
Elson et al. (1998) for inner parts of LMC cluster NGC\,1818.
We expect that these reasonable assumptions for $\alpha$ and $f_{\rm{bin}}$ 
combined with a regular model grid should prevent biases in the 
determination of the parameter space for each cluster.

\section{Statistical tools}
\label{stattools}

The physical parameters of each cluster were determined
by statistical comparisons between synthetic CMDs 
from a grid of models and the observed one.
This method combines a CMD modelling process that has two 
very important qualities: i) it potentially
mimics the effects of photometric uncertainties and unresolved binaries, 
therefore realistically reproducing the observed CMD features; 
ii) it is based on objective criteria to determine
which models best reproduce the data.
This is a robust approach that avoids the subjectivity 
inherent to visual isochrone fits and that is able to reveal any
model solutions that may go undetected in this simpler and more 
popular method.

There are several papers devoted to establishing such statistical tools,
both in the context of CSPs 
(Gallart et al. 1999; Hernandez et al. 1999; Dolphin 2002) and SSPs 
(Valls-Gabaud \& Lastennet 1999; Kerber et al. 2002; Kerber \& Santiago 2005).
Since our data are not very deep, here we prefer to avoid a
two-dimensional comparison method (which uses star counts throughout
the CMD plane). Rather, we follow a more simplistic and appropriate
approach that makes use of both the MS ridge line and the RC position.
This approach is less sensitive to photometric uncertainties and 
incompleteness. 
The simultaneous comparison of these two CMD features ensures
a reliable criterion to establish what the best models are, 
as demonstrated by control experiments.  

\begin{figure}
\resizebox{\hsize}{!}{\includegraphics{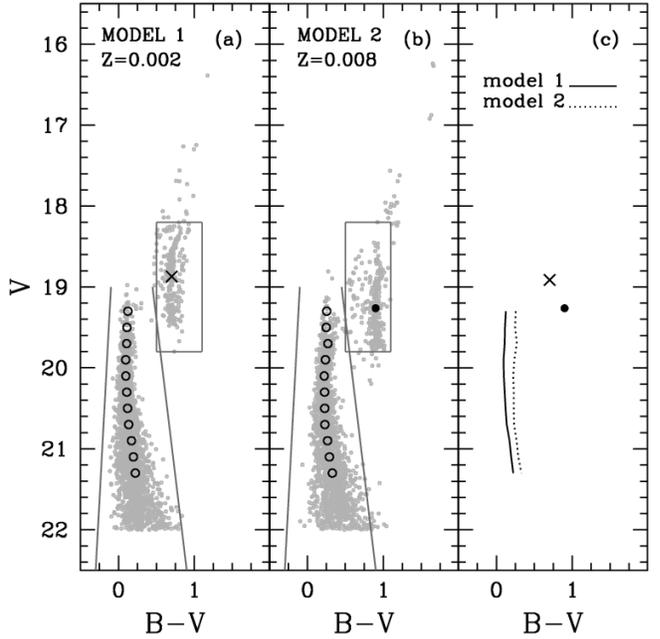}}
\caption{
Synthetic CMDs and their associated MS ridge lines and RC positions.
As labelled in the panels, the models adopt different 
Z values.
Both CMDs were generated using the following parameters:
log($\tau$/yr)=9.00, $(m-M)_{0}=18.50$, E(B-V)=0.05.
}
\label{model_vs_model}
\end{figure}

For each CMD, an MS ridge line was determined using the (B-V) median 
positions at each V magnitude bin along the MS.
Fig. \ref{model_vs_model}ab presents two synthetic CMDs 
and their MS fiducial lines computed in this way. They can be visually 
compared in panel c.
To minimize contamination by spurious objects (certainly present in 
the observed CMDs) 
and stars belonging to other CMD branches, only stars inside the two lines
as shown in this figure were considered as part of the MS.

The $\chi^{2}$ statistic was employed to compare the 
model (mod) and data (obs) colours, being computed for the $N_{\rm{bin}}$
magnitude bins along the MS according to the expression

$$\chi^{2} = \frac{1}{N_{\rm{bin}}-1}\sum_{i=1}^{N_{\rm{bin}}}
\left[\frac{(B-V)_{\rm{obs},i}-(B-V)_{\rm{mod},i}}{\sigma_{\rm{B-V,MS}}}\right]^{2} ~,$$

where $\sigma_{\rm{B-V,MS}}$ is the dispersion in the median colour position
for the i-th V magnitude bin in the model (typically $\sim 0.01$ mag, 
as determined by 
control experiments).

The RC position was determined by using the median position in the CMD plane 
of the stars that likely belong to this phase and that fall inside an
appropriately chosen CMD box.
Therefore the (V,B-V) RC coordinates were determined by the median in the 
V magnitude and colour distributions, respectively. 
This process is also depicted in Fig. \ref{model_vs_model}. 

To compare the RC positions, we define a distance on the CMD plane,
given by 

$$\delta_{RC} = \sqrt{[\frac{V_{\rm{obs}} - 
V_{\rm{mod}}}{\sigma_{\rm{V,RC}}}]^{2}
+ [\frac{(B-V)_{\rm{obs}} - (B-V)_{\rm{mod}}}{\sigma_{\rm{B-V,RC}}}]^{2}} $$

where $\sigma_{\rm{V,RC}}$ and $\sigma_{\rm{B-V,RC}}$ are the dispersions in 
V and (B-V) coordinates for the model RC median position ($\sim 0.03$ and
$\sim 0.01$, respectively). 

We considered as the best models those that \emph{simultaneously} 
satisfy the following criteria:

$$\chi^{2} \le \chi^{2}_{\rm{min}} + n\sigma_{\chi}$$

$$\delta_{RC} \le \delta_{\rm{RC,min}} + n\sigma_{\delta}~,$$

where the index ``min'' refers to the model with the minimum value of each
statistic, and $\sigma_{\chi}$ and $\sigma_{\delta}$ are respectively
the expected dispersions in the distribution of $\chi^{2}$ and
$\delta$ values. They were determined by comparing synthetic CMDs 
of the same model (both dispersions are usually $\sim$ 1.0 to 2.0). 
The parameter ``n'' is the necessary integer number of $\sigma$ in
each statistic 
so that at least three models satisfy the criteria given above.
As shown in Sect. \ref{results}, $n$ is typically 2 or 3. However, in some 
cases, solutions were found only with higher values of $n$, 
indicating less reliable solutions. 
The two statistics have the same weight in the determination of 
the best models, meaning that the MS and the RC are equally 
important in the final solution for each cluster.

To test our statistical approach and to quantify the formal 
uncertainties associated with it we ran some control experiments 
where synthetic CMDs were used as ``observed CMDs'' with known input 
parameters.
The results from these experiments are detailed in the appendix.
As expected, for each experiment the best models recovered by our 
statistical tools have similar parameters when compared to the model 
used to create the ``observation''.

The value of each physical parameter derived for a cluster  
was assumed as the mean of its best models, 
the associated uncertainty being the maximum value among the following ones:
i) the dispersion calculated for the best models;
ii) the formal uncertainty, as determined by a control experiment
(see Appendix);
iii) the half bin size in the model grid.

\section{Results}
\label{results}

The resulting physical parameters of each cluster are presented in 
Table \ref{parameters}, including the conversion of 
log($\tau$) and Z to $\tau$ and [Fe/H] values
to help with future comparisons and use of these parameters.
In the last column the parameter $n$ is presented to reflect the 
reliability of the result. 

\begin{table*}
\caption{Physical parameters derived for all clusters}
\label{parameters}
\centering
\begin{tabular}{l c c c c c c c}
\hline\hline
Cluster & log($\tau$/yr) & $\tau$/Gyr & $Z$ & [Fe/H] & $(m-M)_{0}$ & $E(B-V)$ & n ($\sigma$)\\
\hline
NGC\,1651 & $9.30 \pm 0.03$ & $2.00 \pm 0.15$ & $0.004 \pm 0.001$ & $-0.70 \pm 0.10$ & $18.53 \pm 0.03$ & $0.11 \pm 0.01$ & 2 \\
NGC\,1718 & $9.31 \pm 0.03$ & $2.05 \pm 0.15$ & $0.008^{+0.002}_{-0.001}$ & $-0.40 \pm 0.10$ &  $18.73 \pm 0.07$ & $0.10 \pm 0.03$ & 3 \\
NGC\,1777 & $9.06 \pm 0.04$ & $1.15 \pm 0.15$ & $0.005 \pm 0.001$ & $-0.60 \pm 0.10$ & $18.56 \pm 0.07$ & $0.10 \pm 0.03$ & 4 \\
NGC\,1831 & $8.85 \pm 0.05$ & $0.70 \pm 0.10$ & $0.016 \pm 0.003$ & $-0.10\pm0.10$ & $18.23 \pm 0.09$ & $0.01 \pm 0.02$ & 7 \\
NGC\,1856 & $8.47 \pm 0.04$ & $0.30 \pm 0.25$ & $0.008^{+0.002}_{-0.001}$ & $-0.40\pm0.10$ & $18.37 \pm 0.10$ & $0.21 \pm 0.02$ & 2 \\
NGC\,1868 & $9.05 \pm 0.03$ & $1.10 \pm 0.10$ & $0.004 \pm 0.001$ & $-0.70\pm0.10$ & $18.33 \pm 0.06$ & $0.04 \pm 0.01$ & 3 \\
NGC\,2121 & $9.46 \pm 0.07$ & $2.90 \pm 0.50$ & $0.008^{+0.002}_{-0.001}$ & $-0.40\pm0.10$ & $18.24 \pm 0.04$ & $0.07 \pm 0.02$ & 5 \\
NGC\,2155 & $9.48 \pm 0.03$ & $3.00 \pm 0.25$ & $0.004 \pm 0.001$ & $-0.70\pm0.10$ & $18.32 \pm 0.04$ & $0.02 \pm 0.01$ & 2 \\
NGC\,2162 & $9.10 \pm 0.03$ & $1.25 \pm 0.10$ & $0.008 \pm 0.002$ & $-0.40 \pm0.10$ & $18.35 \pm 0.08$ & $0.03 \pm 0.02$ & 3 \\
NGC\,2173 & $9.21 \pm 0.04$ & $1.60 \pm 0.20$ & $0.005 \pm 0.001$ & $-0.60\pm0.10$ & $18.58 \pm 0.12$ & $0.07 \pm 0.02$ & 2 \\
NGC\,2209 & $9.08 \pm 0.03$ & $1.20 \pm 0.10$ & $0.006 \pm 0.001$ & $-0.50\pm0.10$ & $18.43 \pm 0.09$ & $0.15 \pm 0.03$ & 4 \\
NGC\,2213 & $9.23 \pm 0.04$ & $1.70 \pm 0.20$ & $0.004 \pm 0.001$ & $-0.70\pm0.10$ & $18.56 \pm 0.08$ & $0.06 \pm 0.03$ & 2 \\
NGC\,2249 & $9.00 \pm 0.03$ & $1.00 \pm 0.10$ & $0.007 \pm 0.001$ & $-0.45\pm0.10$ & $18.27 \pm 0.06$ & $0.01 \pm 0.02$ & 2 \\
SL\,506   & $9.35 \pm 0.03$ & $2.25 \pm 0.15$ & $0.007 \pm 0.001$ & $-0.45\pm0.10$ & $18.48 \pm 0.06$ & $0.08 \pm 0.03$ & 2 \\
SL\,663   & $9.50 \pm 0.05$ & $3.15 \pm 0.40$ & $0.004 \pm 0.001$ & $-0.70\pm0.10$ & $18.32 \pm 0.07$ & $0.07 \pm 0.02$ & 3 \\
\hline
\end{tabular}
\end{table*} 

In Figs \ref{n1856_data_vs_model}-\ref{n2121_data_vs_model} we compare
the observed and synthetic CMDs. The sequence of presentation is the same
as in Fig. \ref{allcmds} to underline the evolutionary sequence for 
these clusters. 
In all figures, panel {\bf a} presents the data and panel {\bf b} shows 
a synthetic CMD generated from one of the best models according to the
criteria discussed in the previous section.
The corresponding fiducial lines and RC positions are compared in 
panel {\bf c}.  
For 11 clusters we have $n \le 3$ and, in general, the models 
reproduce well both the MS fiducial line, the
RC position and the spread in magnitude and colour. 
However, in a few cases there are some discrepancies. They
will be commented on below together with the cases where
$n > 3$.
In the next subsections we will also compare our age and metallicity 
results with the ones found in the literature summarized in 
Table \ref{lit_age_FeH}. 

In terms of distance modulus and reddening there is a lack direct 
determinations for a significant number of clusters in our sample. 
The only recent exception to this rule is the work by
McLaughlin \& van der Marel (2005), who compared SSP model predictions 
with integrated colours. 
As uncertainties based on integrated cluster light tend to be larger,
we prefer to use the results of McLaughlin \& van der Marel (2005)
to assess the extent to which their method reproduces the results of
resolved photometry rather than to assess the reliability of our
method itself.
Typical LMC values adopted for E(B-V) and reddening 
are $\sim 0.07$ (Burstein \& Heiles 1982; Schlegel et al. 1998) 
and $\sim 18.50$ (see Clementini et al. 2003 for a review about
the distance to the LMC).
Thus, in the next section we will comment only on results that deviate
strongly from these fiducial values.  

\subsection{Cluster-by-cluster}
\label{clusterbycluster}

\noindent
{\bf NGC\,1856}

This is the youngest cluster and the nearest to the LMC optical centre
in our sample, which explains the high reddening value obtained, 
in accordance with the ones recovered by McLaughlin \& van der Marel (2005) 
(typically $E(B-V) \simeq 0.20\pm0.05$). 
Notice that the synthetic CMD reproduces well both the MS fiducial line and 
RC features.
However, the model did not reproduce the stars located beyond the
upper limit of the MS, which are likely binaries and/or blue stragglers.

Our age result is consistent with the upper limit of Elson \& Fall
(1988) and with the one obtained by Girardi et al. (1995).
It is also consistent with the lowest age solution from 
Beasley et al. (2002).
However these authors recovered an almost solar metallicity,
higher than the value inferred here.
As expected from the age-metallicity degeneracy, their highest age 
solution has a lower value for metallicity, 
compatible with our value within the uncertainties.

\begin{figure}
\resizebox{\hsize}{!}{\includegraphics{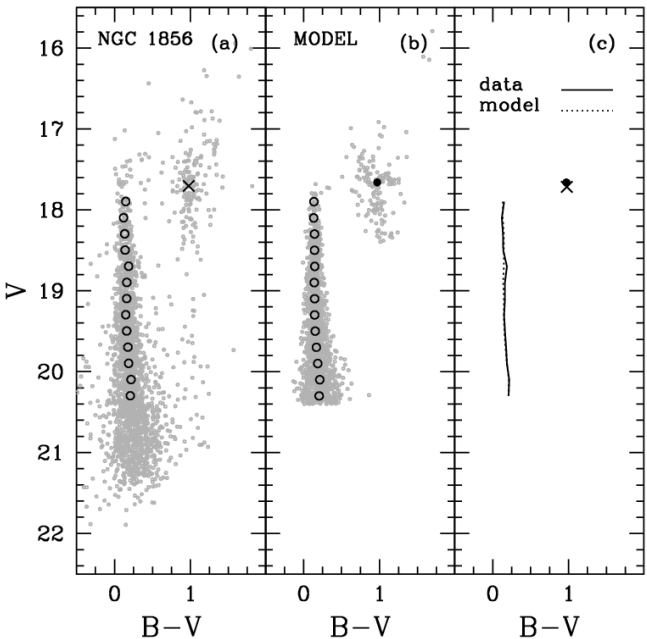}}
\caption{Data (panel {\bf a}) vs. model (panel {\bf b}) 
comparison for NGC\,1856. 
Panel {\bf c} confronts the fiducial lines and RC positions traced
by the points shown in panel {\bf a} and {\bf b}.
The synthetic CMD was generated using the following
parameters: log($\tau$/yr)=8.45, $Z$=0.008, $(m-M)_{0}=18.45$, E(B-V)=0.21.
}
\label{n1856_data_vs_model}
\end{figure}

\smallskip

\noindent
{\bf NGC\,1831}

Only solutions with n=7, i.e., the best 3 models spread out
in a region $7 \times \sigma$ away from the minimum $\chi^2$, 
were found for the second youngest cluster in our sample.
The difficulty here was to find models with the appropriate color 
difference between the MS fiducial line and RC position.
The best models still have bluer MS fiducial lines and redder 
RC positions than the data.
On the other hand, these solutions visually mimic the stars at the
bright end of the MS and the RC spread.  

Our results confirm a high metallicity value, also
found by OSSH (using only one star) and Kerber \& Santiago (2005),
although this latter work recovered a slightly younger age than found here.
The upper age limits obtained by Elson \& Fall (1988) and 
Leonardi \& Rose (2003) are consistent
with our result, although the metallicity recovered by the latter 
is significantly lower.

As determined by Kerber \& Santiago (2005), we found a low reddening value,
but they derived a larger distance modulus (18.70$\pm$0.03) than
obtained here. 
This and other discrepancies between the results in this work
and those obtained by Kerber \& Santiago (2005) are likely due to
two main reasons:
i) here we analyse not only the MS, but also the RC, 
a very important phase which, when combined with the MS, better constrains the
best models (see Appendix for control experiments);
ii) the systematic biases found in the photometry in both studies were not 
corrected by the same method, likely leading to some photometric 
discrepancies.

\begin{figure}
\resizebox{\hsize}{!}{\includegraphics{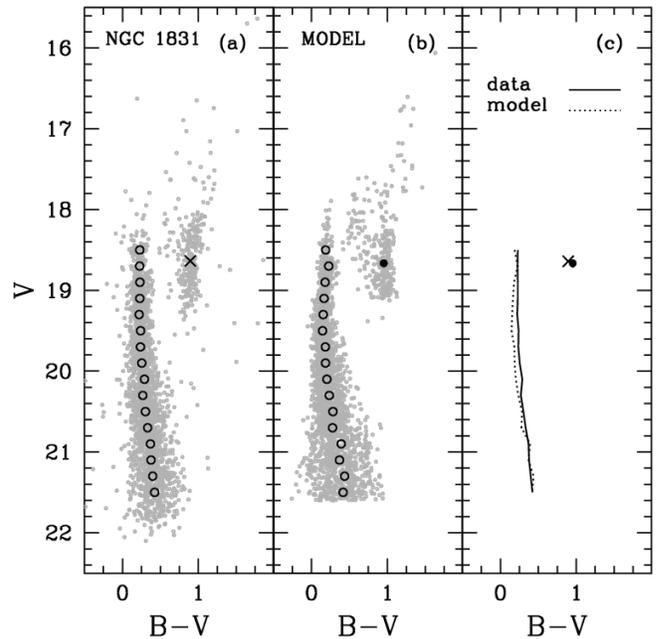}}
\caption{The same as in Fig. \ref{n1856_data_vs_model} but for NGC\,1831.
The synthetic CMD was generated using the following
parameters: log($\tau$/yr)=8.85, $Z$=0.016, $(m-M)_{0}=18.20$, E(B-V)=0.00.
}
\label{n1831_data_vs_model}
\end{figure}

\smallskip

\noindent
{\bf NGC\,2249}

The agreement between data and model is close for this cluster.
Our recovered age is only in agreement with Elson \& Fall (1988), 
the other previous values being significantly younger.
Our metallicity result is in accordance with those by
Leonardi \& Rose (2003) and Mackey \& Gilmore (2003).
We reach low reddening and distance modulus values. As
most of the extinction is internal to the LMC, clusters seen in
the foreground should have lower reddening.

\begin{figure}
\resizebox{\hsize}{!}{\includegraphics{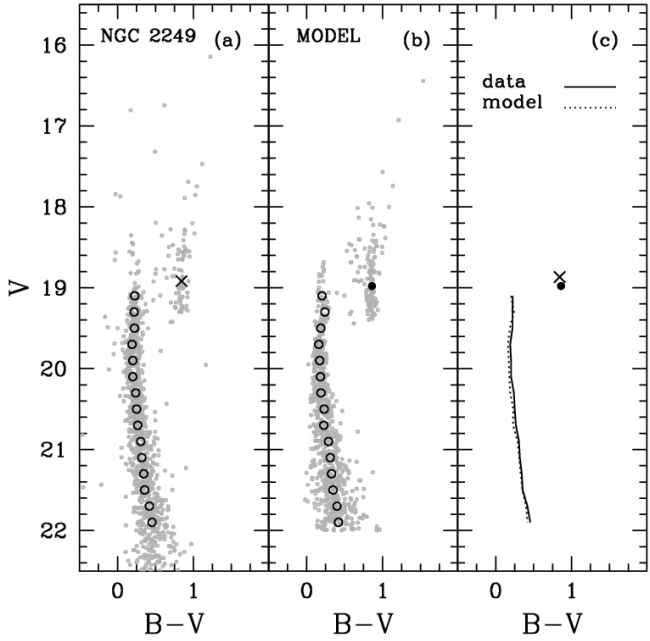}}
\caption{The same as in Fig. \ref{n1856_data_vs_model} but for NGC\,2249.
The synthetic CMD was generated using the following
parameters: log($\tau$/yr)=9.00, $Z$=0.008, $(m-M)_{0}=18.25$, E(B-V)=0.00.
}
\label{n2249_data_vs_model}
\end{figure}

\smallskip

\noindent
{\bf NGC\,1868}

Again the best models reproduced well the MS fiducial line and the RC
 characteristics. 

The ages found in the literature are systematically
younger than ours, although the upper limits of 
Elson \& Fall (1988), Leonardi \& Rose (2003) and Kerber \& Santiago (2005)
are consistent with our determination.
The metallicity we obtain is consistent with those from OSSH and 
Leonardi \& Rose (2003) (although this latter is highly uncertain), 
but it is lower than the one recovered by Kerber \& Santiago (2005).

As in NGC\,1831, we obtained a low reddening value, as was
found by Kerber \& Santiago (2005), but again our best models 
have a significantly smaller distance modulus.

\begin{figure}
\resizebox{\hsize}{!}{\includegraphics{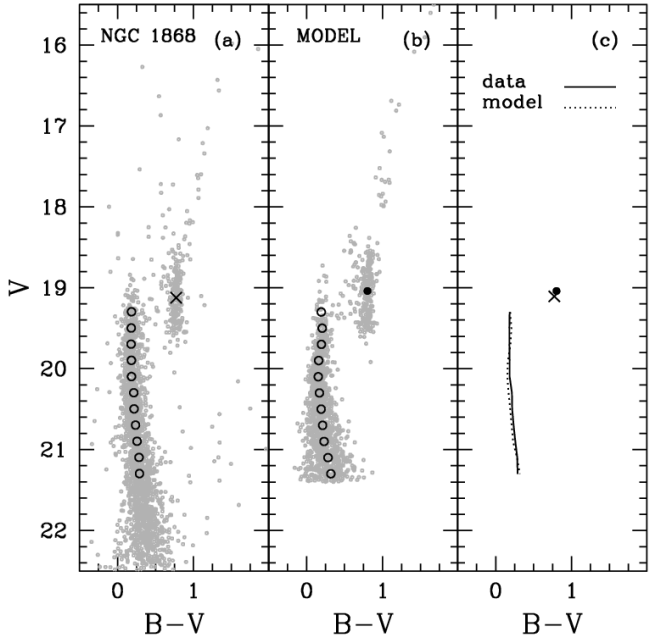}}
\caption{The same as in Fig. \ref{n1856_data_vs_model} but for NGC\,1868.
The synthetic CMD was generated using the following
parameters: log($\tau$/yr)=9.05, $Z$=0.004, $(m-M)_{0}=18.35$, E(B-V)=0.04.
}
\label{n1868_data_vs_model}
\end{figure}

\smallskip

\noindent
{\bf NGC\,2162}

The best solutions correctly fit the RC position, but have MS fiducial lines
slightly bluer than the data. These solutions are limited by $n$=3. 
The recovered age matches the results by Geisler et al. (1997) and 
the upper limit of Girardi et al. (1995). 
Our inferred metallicity agrees with OSSH within the 
uncertainties. 
The solution found by Leonardi \& Rose (2003) is 
significantly older and more metal-poor than most other results.
Low reddening and distance modulus values were found for NGC\,2162,
again consistent with a foreground location relative to most LMC stars.
%McLaughlin \& van der Marel (2005) obtained typically E(B-V)=$0.03\pm0.08$. 

\begin{figure}
\resizebox{\hsize}{!}{\includegraphics{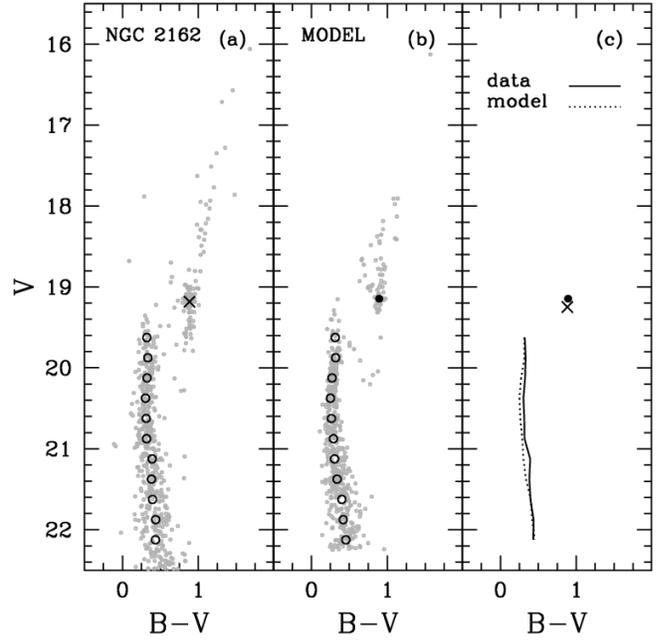}}
\caption{The same as in Fig. \ref{n1856_data_vs_model} but for NGC\,2162.
The synthetic CMD was generated using the following
parameters: log($\tau$/yr)=9.10, $Z$=0.008, $(m-M)_{0}=18.35$, E(B-V)=0.03.
}
\label{n2162_data_vs_model}
\end{figure}

\smallskip

\noindent
{\bf NGC\,1777}

Although the best solutions require $n = 4$, they visually reproduce all 
the observed features in the CMD. The age result is in good agreement 
with previous determinations.
Our recovered  metallicity is consistent with the lower limit 
determined by OSSH and by Leonardi \& Rose (2003), if one admits a more 
typical (and likely more realistic) uncertainty of $\sim 0.10$ for 
the value from the latter. 

\begin{figure}
\resizebox{\hsize}{!}{\includegraphics{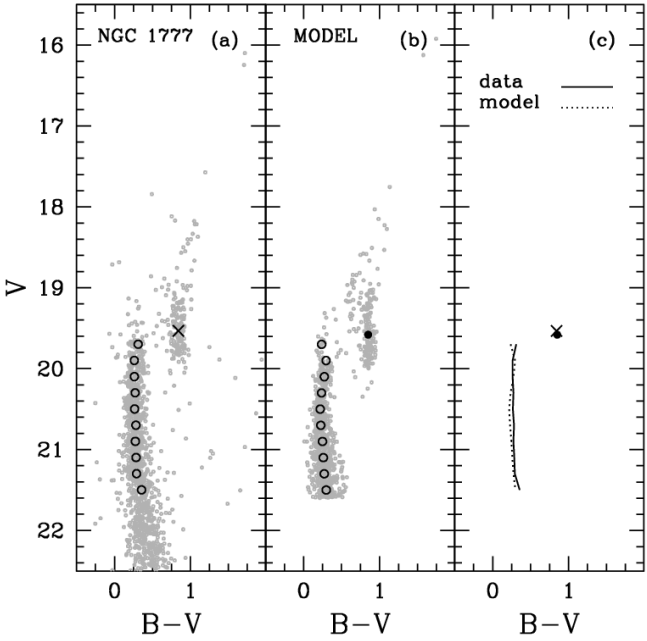}}
\caption{The same as in Fig. \ref{n1856_data_vs_model} but for NGC\,1777.
The synthetic CMD was generated using the following
parameters: log($\tau$/yr)=9.05, $Z$=0.004, $(m-M)_{0}=18.55$, E(B-V)=0.11.
}
\label{n1777_data_vs_model}
\end{figure}

\smallskip

\noindent
{\bf NGC\,2209}

The best models correctly reproduce the RC features and nicely mimic 
the unresolved binaries in the MS termination. The n=4$\sigma$ level 
of these solutions is related to the difficulty in reproducing the MS 
fiducial line, since the models are slightly but systematically 
bluer than the data. 

The ages found in the literature agree with our
result. Unfortunately we did not find a unique published metallicity 
determination for this cluster, but the crude estimate done by 
Mackey \& Gilmore (2003) closely agrees with our result, meaning that 
this cluster has a typical metallicity for an IAC.

A high reddening value was determined for this cluster, 
although it is located at a distance modulus slightly lower than the typical 
LMC value ($(m-M)_0$=18.50).
McLaughlin \& van der Marel (2005) derived E(B-V) values that are 
typically half of what we infer here,
but with large uncertainties ($\sim 0.10$).

\begin{figure}
\resizebox{\hsize}{!}{\includegraphics{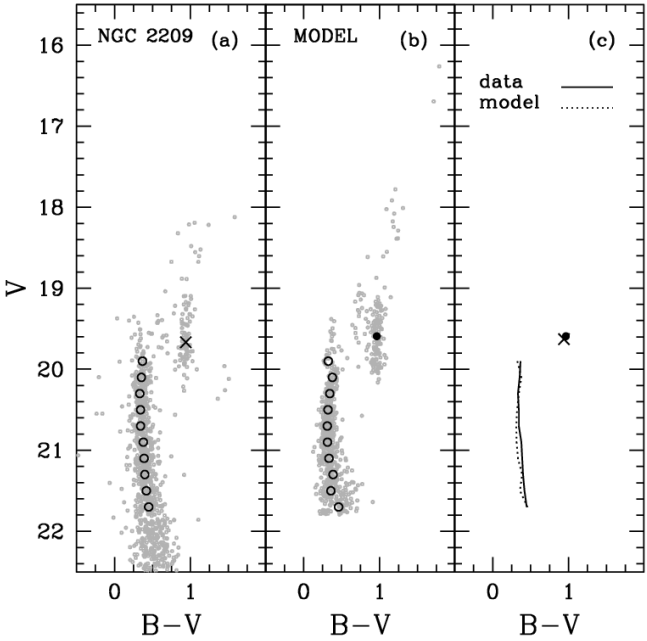}}
\caption{The same as in Fig. \ref{n1856_data_vs_model} but for NGC\,2209.
The synthetic CMD was generated using the following
parameters: log($\tau$/yr)=9.05, $Z$=0.006, $(m-M)_{0}=18.40$, E(B-V)=0.16.
}
\label{n2209_data_vs_model}
\end{figure}

\smallskip

\noindent
{\bf NGC\,2213}

The agreement between model and data CMDs is very close for this cluster,
including the RGB and the onset of the SGB.
Our age estimate also agrees well with that of Geisler et al. (1997),
but it is lower than that from Leonardi \& Rose (2003) and higher than
the one from Girardi \& Bertelli (1998). 
If one accepts a more typical uncertainty of $\sim 0.10$ in 
log($\tau$) than that quoted by Leonardi \& Rose (2003), their value 
becomes compatible with our one. 
The metallicity result from OSSH is much higher than our value, but
this spectroscopic result was based on only one star. 

\begin{figure}
\resizebox{\hsize}{!}{\includegraphics{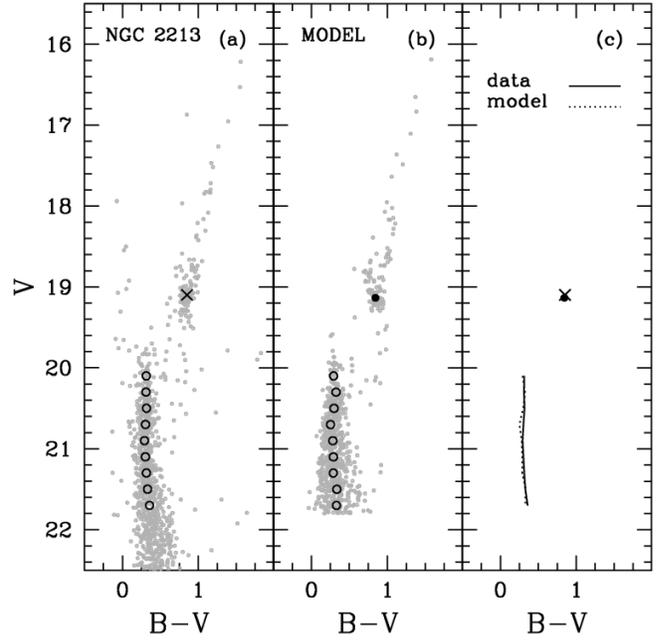}}
\caption{The same as in Fig. \ref{n1856_data_vs_model} but for NGC\,2213.
The synthetic CMD was generated using the following
parameters: log($\tau$/yr)=9.20, $Z$=0.004, $(m-M)_{0}=18.55$, E(B-V)=0.06.
}
\label{n2213_data_vs_model}
\end{figure}

\smallskip

\noindent
{\bf NGC\,2173}

This is another example of a quality fit between data and model, which
also adequately reproduces the observed SGB and RGB. 
The only weak point seems to be the RC spread, lower in the model than
in the data.

A large number of results are found in the literature. 
Our age determination is consistent with Geisler et al. (1997) and with 
the recent works done by Bertelli et al. (2003) 
(although they argued that this cluster had a prolonged star formation)
and Woo et al. (2003),
being higher than the one derived by Girardi et al. (1995) and 
significantly lower than the one obtained by Leonardi \& Rose (2003).
However, if the result from Leonardi \& Rose (2003) is correct, 
then this cluster would be the second one located in the age gap. 
Also, these authors find a very low metallicity compared to a typical IAC.
On the other hand, the higher metallicity value from OSSH 
is not confirmed by the two independent solutions found by 
Woo et al. (2003) and Bertelli et al. (2003), which are based on
VLT/CMDs and which use Y$^{2}$ and Padova isochrones, respectively.

The distance modulus and reddening for this cluster 
were also derived by Bertelli et al. (2003) and Woo et al. (2003).
While the latter found results consistent with ours,
the former obtained a distance modulus $\sim 0.10$ lower than 
the one found by us; however, their acceptable solutions,
based on different criteria, presented some internal discrepancies 
in the reddening value (and in the metallicity), their intermediate 
reddening solution (with Z=0.003) being in accordance with ours. 

\begin{figure}
\resizebox{\hsize}{!}{\includegraphics{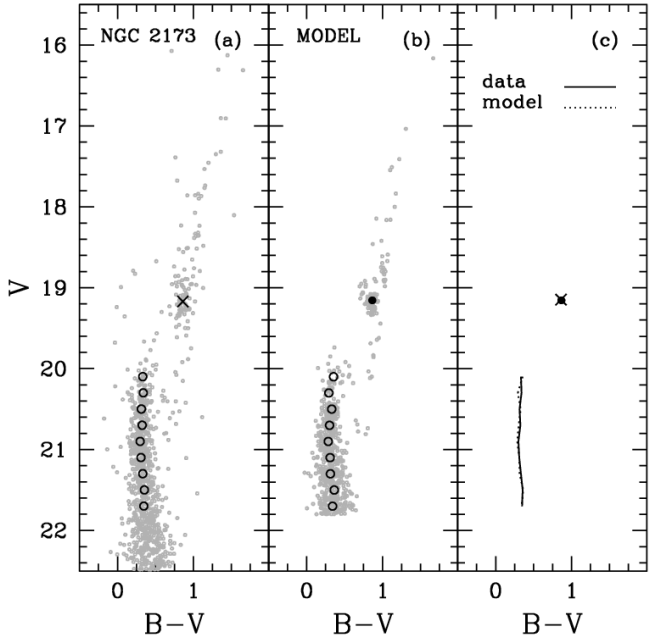}}
\caption{The same as in Fig. \ref{n1856_data_vs_model} but for NGC\,2173.
The synthetic CMD was generated using the following
parameters: log($\tau$/yr)=9.20, $Z$=0.004, $(m-M)_{0}=18.60$, E(B-V)=0.07.
}
\label{n2173_data_vs_model}
\end{figure}

\smallskip

\noindent
{\bf NGC\,1651}

Both the MS fiducial line and the RC features are well reproduced by
the models.
However, the SGB in the synthetic CMD seems to be more scattered
and the RGB more elongated towards the bright end than in the observed CMD.

The age results found in the literature are in accordance with 
each other and with this work.
Taking the uncertainties into account, 
our resulting metallicity agrees with the previous 
metal-poor determinations from Dirsch et al. (2000) 
and Leonardi \& Rose (2003).
The result from OSSH is metal-richer than ours, but it
was based on a single star, whereas the metallicity 
derived by Sarajedini et al. (2002) is, surprisingly, 
almost solar. 

Recently the distance modulus was determined by 
Sarajedini et al. (2002) and Grocholski et al. (2005), 
analysing the RC stars with near infrared data.
They found $(m-M)_{0}=18.55\pm0.12$ and $(m-M)_{0}=18.50\pm0.06$,
respectively, in accordance with our result. 
Notice also that their adopted reddening value, based on Burstein
\& Heiles (1982) and Schlegel et al. (1998), was $E(B-V)=0.12\pm0.02$,
again in agreement with our determination.  

\begin{figure}
\resizebox{\hsize}{!}{\includegraphics{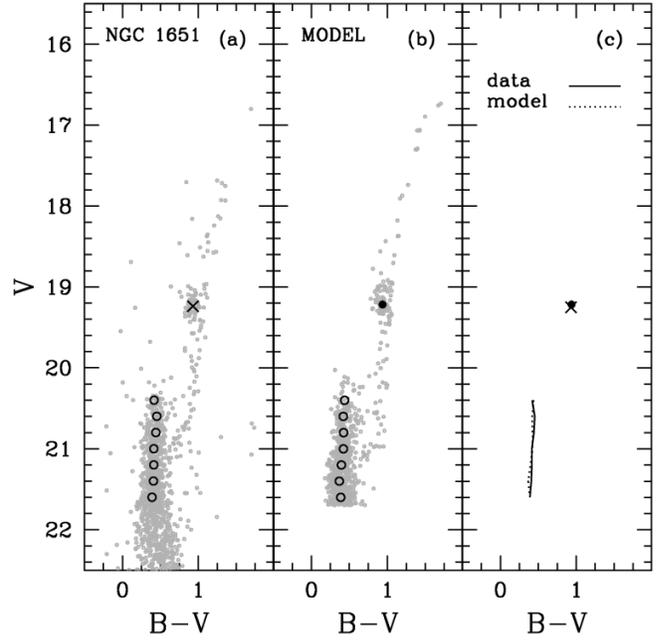}}
\caption{The same as in Fig. \ref{n1856_data_vs_model} but for NGC\,1651.
The synthetic CMD was generated using the following
parameters: log($\tau$/yr)=9.30, $Z$=0.004, $(m-M)_{0}=18.50$, E(B-V)=0.11.
}
\label{n1651_data_vs_model}
\end{figure}

\smallskip

\noindent
{\bf NGC 1718}

Although the MS fiducial line has been well fitted by the best models, 
the model RC is slightly brighter than the data. 
Notice also the presence of stars near but above the MS termination,
not reproduced by the models, that are probable unresolved binaries.
Another group of stars well above the MSTO may be field stars,
possibly belonging to horizontal branch.
The best models still visually reproduce very well both the SGB and
RGB, supporting our solutions.

Our determinations for age agree with the results from
Elson \& Fall (1988) and with the lowest age solution found by 
Beasley et al. (2002). 
However, only the upper limit predicted for metallicity by these latter 
authors is compatible with our value. 
As expected, this discrepancy in metallicity becomes greater if we 
compare our result
with that from the highest age solution found by Beasley et al. (2002). 
On the other hand, our derived metallicity is typical of an IAC in the LMC, 
as attested by the metallicity estimate by Mackey \& Gilmore (2003).

We infer for this cluster the highest distance modulus 
in the sample and a moderately high value of reddening. 
Although less accurate, the reddening results from 
McLaughlin \& van der Marel (2005) ($\sim 0.10\pm0.10$) point in the
same direction.

\begin{figure}
\resizebox{\hsize}{!}{\includegraphics{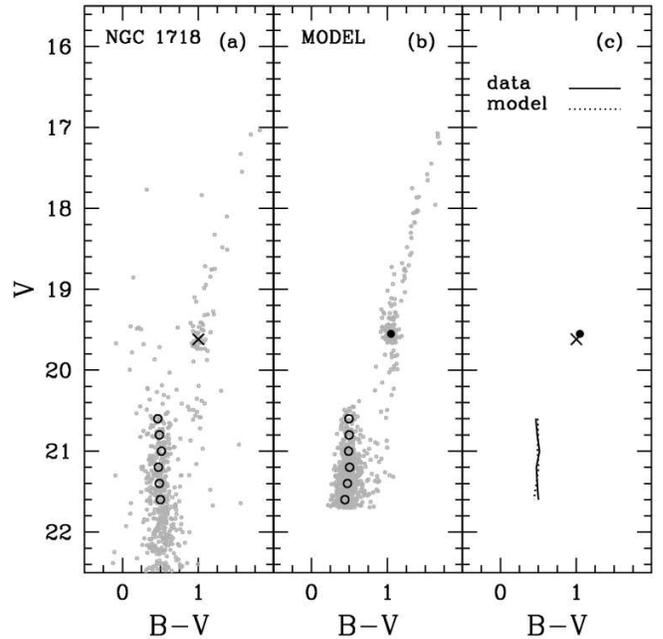}}
\caption{The same as in Fig. \ref{n1856_data_vs_model} but for NGC\,1718.
The synthetic CMD was generated using the following
parameters: log($\tau$/yr)=9.30, $Z$=0.008, $(m-M)_{0}=18.70$, E(B-V)=0.10.
}
\label{n1718_data_vs_model}
\end{figure}

\smallskip

\noindent
{\bf SL 506 (Hodge\,14)}

The observed CMD of this cluster is well reproduced by the best models
in all its features. 
Although the two previous results from the literature have slightly
younger ages, they are still consistent.
Our determination also agrees very well with metallicity values of 
Kerber \& Santiago (2005) and are consistent with OSSH, who found a 
lower value. 

In terms of distance modulus, Kerber \& Santiago (2005) reached almost 
the same result obtained here. 
Their reddening value (E(B-V)=$0.02\pm0.02$)
is marginally consistent with ours, taking into account the
uncertainties in the two determinations.

\begin{figure}
\resizebox{\hsize}{!}{\includegraphics{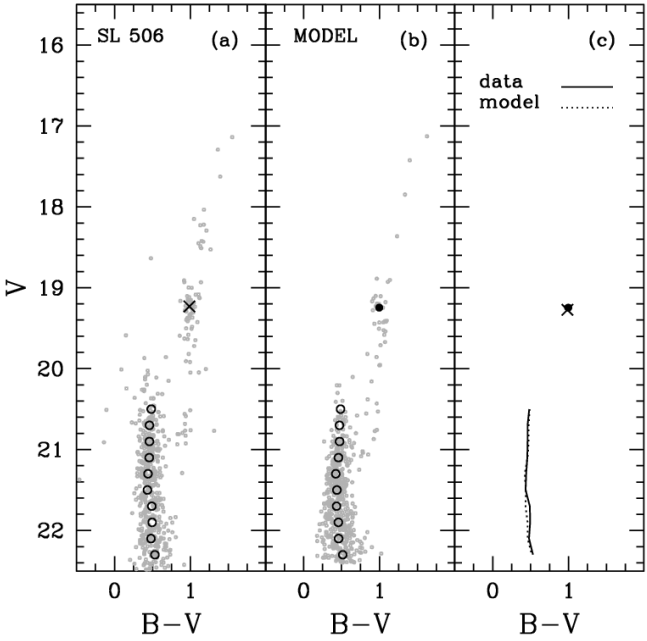}}
\caption{The same as in Fig. \ref{n1856_data_vs_model} but for SL\,506.
The synthetic CMD was generated using the following
parameters: log($\tau$/yr)=9.35, $Z$=0.008, $(m-M)_{0}=18.50$, E(B-V)=0.05.
}
\label{sl506_data_vs_model}
\end{figure}

\smallskip

\noindent
{\bf NGC 2155}

Again the best models well reproduce both the MS fiducial line and 
the RC, but the RGB in the synthetic CMDs seems to be narrower 
than in the observed CMD. 
Above the MSTO it is also possible to identify some likely 
unresolved binaries (mimicked by the artificial CMDs) and
field stars ($\ga 0.60$ mag brighter than the MS termination).
This cluster has a large number of age and metallicity determinations,
and our results are in agreement with them.

We obtained a clear self-consistent result for the distance modulus 
and reddening, both being low values, again indicating a foreground location. 
This result is in agreement with Bertelli et al. (2003),
who obtained their best solutions with $(m-M)_{0}=18.36$ 
and $E(B-V)\sim 0.015-0.028$.
On the other hand, Woo et al. (2003) derived a distance modulus typical
of the LMC (18.50), 0.20 higher than our value, but they also 
recovered a low reddening value ($E(B-V)\sim0.04$).
This low reddening value seems to be confirmed by  
the Burstein \& Heiles (1982) maps, which
indicate E(B-V)=0.03 for this region in the sky.

\begin{figure}
\resizebox{\hsize}{!}{\includegraphics{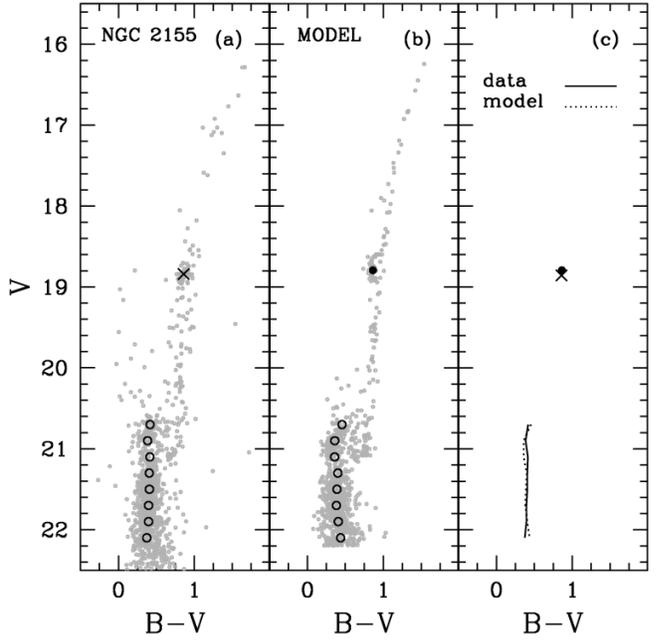}}
\caption{The same as in Fig. \ref{n1856_data_vs_model} but for NGC\,2155.
The synthetic CMD was generated using the following
parameters: log($\tau$/yr)=9.50, $Z$=0.004, $(m-M)_{0}=18.30$, E(B-V)=0.02.
}
\label{n2155_data_vs_model}
\end{figure}

\smallskip

\noindent
{\bf SL 663}

The observed MS fiducial line and RC features are well fitted by the
best models. 
As for NGC 2155, there are some likely field stars and unresolved binaries 
beyond the MSTO, these latter also being reproduced by the artificial CMDs.
The less populated RGB of this cluster is also reproduced by the models.
Our age and metallicity results agree well both with Rich et al. (2001) 
and OSSH. 

\begin{figure}
\resizebox{\hsize}{!}{\includegraphics{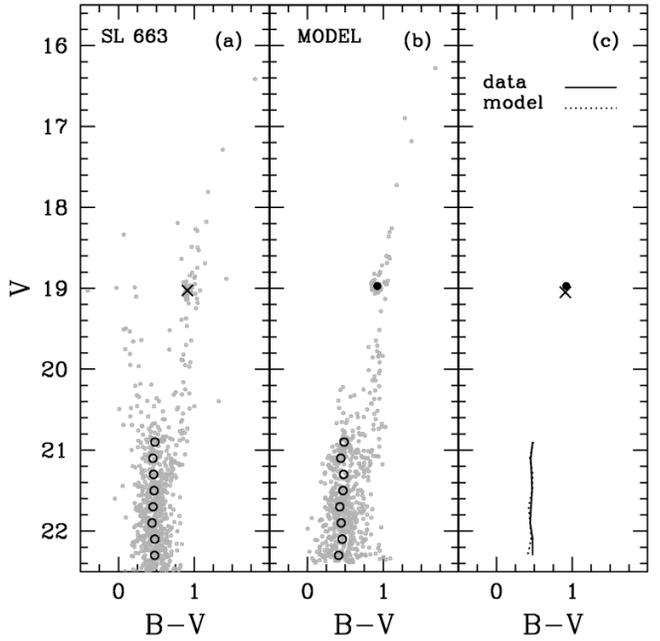}}
\caption{The same as in Fig. \ref{n1856_data_vs_model} but for SL\,663.
The synthetic CMD was generated using the following
parameters: log($\tau$/yr)=9.45, $Z$=0.004, $(m-M)_{0}=18.35$, E(B-V)=0.08.
}
\label{sl663_data_vs_model}
\end{figure}

\smallskip

\noindent
{\bf NGC 2121}

Although they visually reproduce the observed CMD, 
the best models have $n \le 5$. 
As for other previous clusters, the likely field star contamination
and the presence unresolved binaries can be seen.
Since this cluster is more populous, its RGB is correspondingly more 
populated, something that the model CMDs manage to recover.

The age we obtain is in good agreement with 
those found in the literature.
Our inferred metallicity is slightly higher than the 
typical [Fe/H] $\sim -0.6$ value, but it is consistent with it when the
uncertainties are considered. 

According to our results, this cluster has the lowest $(m-M)_0$ in the
sample, but a typical $E(B-V)$ value for the LMC.

\begin{figure}
\resizebox{\hsize}{!}{\includegraphics{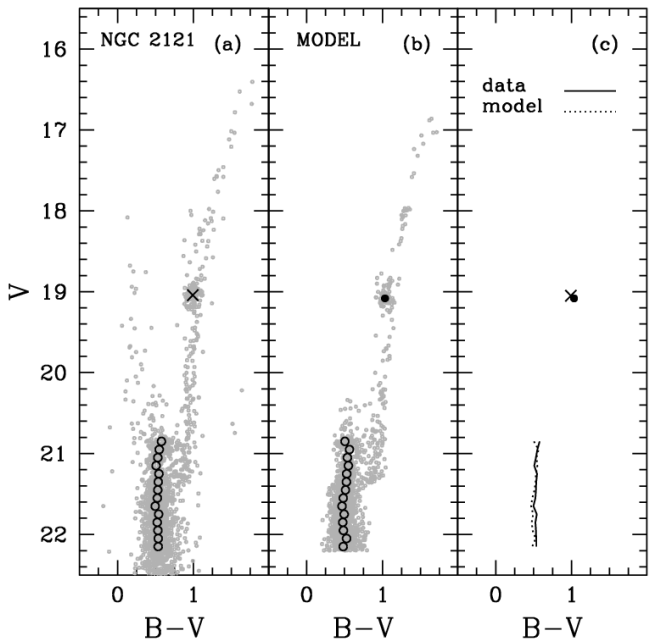}}
\caption{The same as in Fig. \ref{n1856_data_vs_model} but for NGC\,2121.
The synthetic CMD was generated using the following
parameters: log($\tau$/yr)=9.45, $Z$=0.008, $(m-M)_{0}=18.25$, E(B-V)=0.07.
}
\label{n2121_data_vs_model}
\end{figure}

\begin{figure*}
\centering
\includegraphics[width=17cm]{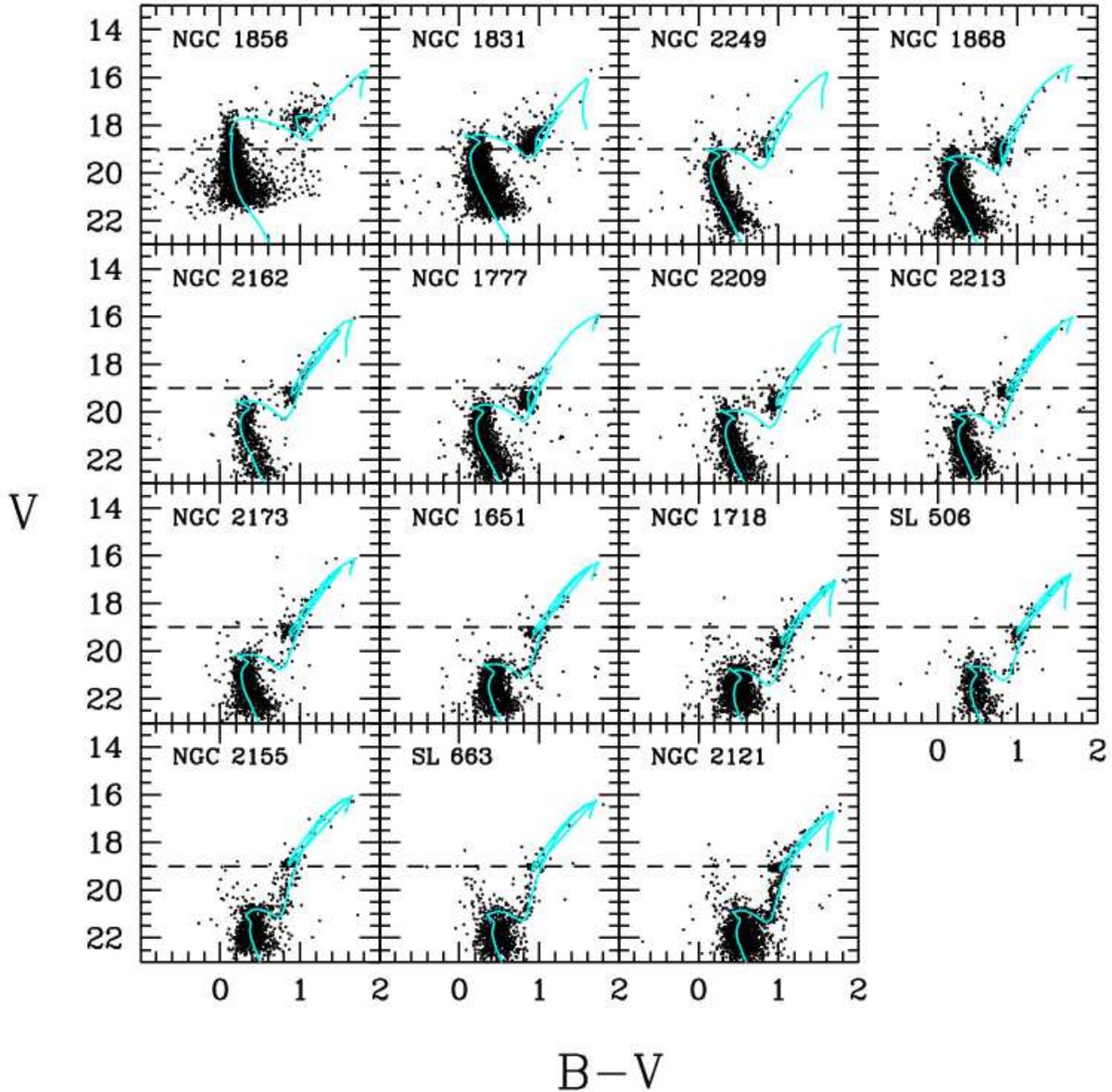}
\caption{
All observed CMDs, as shown in Fig. \ref{allcmds}, but with the 
isochrones corresponding to the best solutions superimposed.}
\label{allcmds_isot}
\end{figure*}

We present the observed CMDs of all clusters, as in Fig. \ref{allcmds}, 
but now with the best model isochrones superposed. 
The isochrones from the best solutions are those presented in 
panels {\bf b} of Figs. \ref{n1856_data_vs_model}-\ref{n2121_data_vs_model}.
This is shown in Fig. \ref{allcmds_isot}, which reveals, 
in general, good isochrone fits to the data, both for MS and RC 
position. 
As expected due to its high $n$ value, the only conspicuous 
exception is NGC\,1831.
In some cases (e.g. NGC\,2249) the isochrone 
is systematically shifted bluewards relative to the bulk of the MS
stars. This shift is expected to compensate for
the effect of unresolved binaries, since these stars 
tend to spread the MS in the redward direction. The effect
is also less prominent for a steeper MS. 
The unresolved binaries are also responsible for 
blurring the bright MS termination, since a pair of equal mass 
stars will reach 0.75 mag brighter than the expected, single-star, 
MS termination magnitude determined by the isochrone line. 

\subsection{The whole cluster sample}
\label{whole}

We compare the properties we infer for the 
whole cluster sample with other previous homogeneous 
determinations found in the literature. 

\begin{figure}
\resizebox{\hsize}{!}{\includegraphics{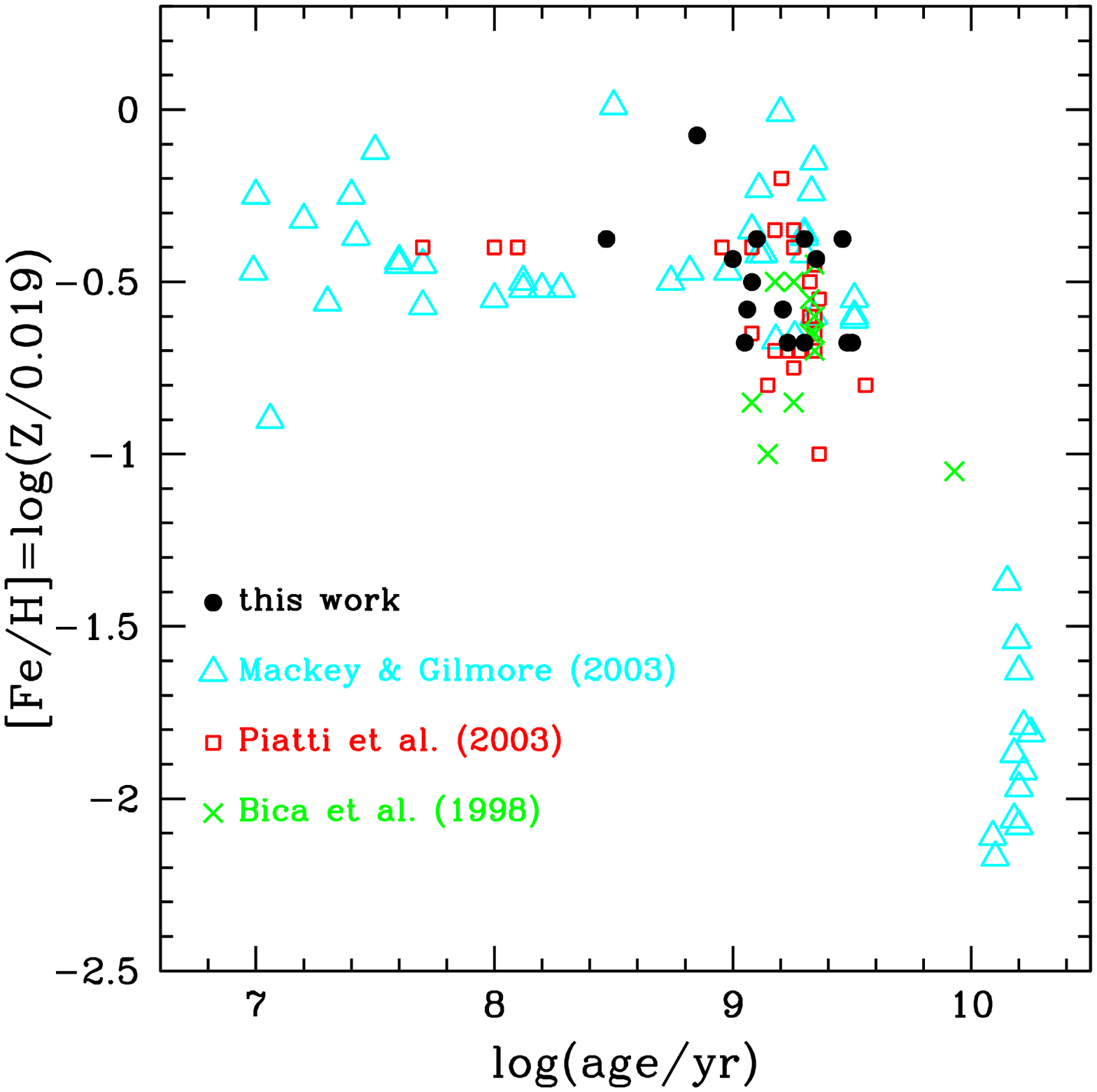}}
\caption{Age-metallicity relation from this work (black points) in 
comparison to the one by Mackey \& Gilmore (2003) 
(open triangles) and the one obtained by Bica et al. (1998) (crosses).}
\label{AMR}
\end{figure}

Fig. \ref{AMR} presents the age-metallicity relation (AMR) for a 
series of works devoted to LMC clusters 
(Bica et al. 1998; Mackey \& Gilmore 2003; Piatti et al. 2003b)
spanning a wide age and metallicities ranges.
Our results for the 15 IACs are also plotted in this figure. 
This reveals the LMC chemical enrichment:
the oldest clusters have the lowest metallicities values
(reaching [Fe/H] $\sim-2.3$), whereas the clusters younger than 
log($\tau$/yr)$\la 9.5$ ($\tau \la 3$ Gyr) are
significantly more metal-rich, belonging to an approximated 
``plateau'' of [Fe/H]$\sim-0.5$, but with a considerable scatter.
The only known cluster in the so-called ``age gap'' range 
($9.5 \la$ log($\tau$) $\la 10.0$ or 3.0 $\la \tau \la 10$ Gyr) is
ESO121-SC03 (marked by a cross, as determined by Bica et al. 1998).
This gap can also be considered as a ``metallicity gap''
for clusters within the range $-1.3 \la$ [Fe/H] $\la -1.0$.

\begin{figure}
\resizebox{\hsize}{!}{\includegraphics{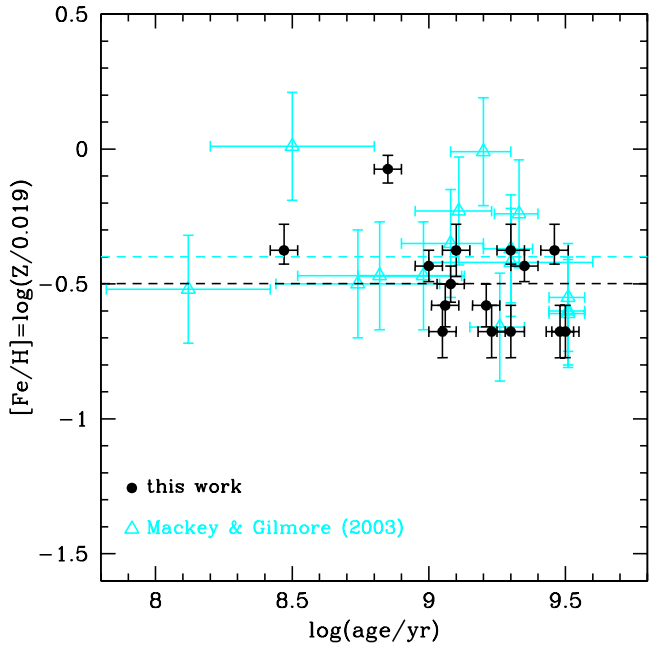}}
\caption{Age-metallicity relation from this work (black) in 
comparison to the one by Mackey \& Gilmore (2003)
(gray) for the same clusters. The dotted (dashed) line represents 
our (their) mean values.}
\label{KvsMG03_AMR}
\end{figure}

\begin{figure}
\resizebox{\hsize}{!}{\includegraphics{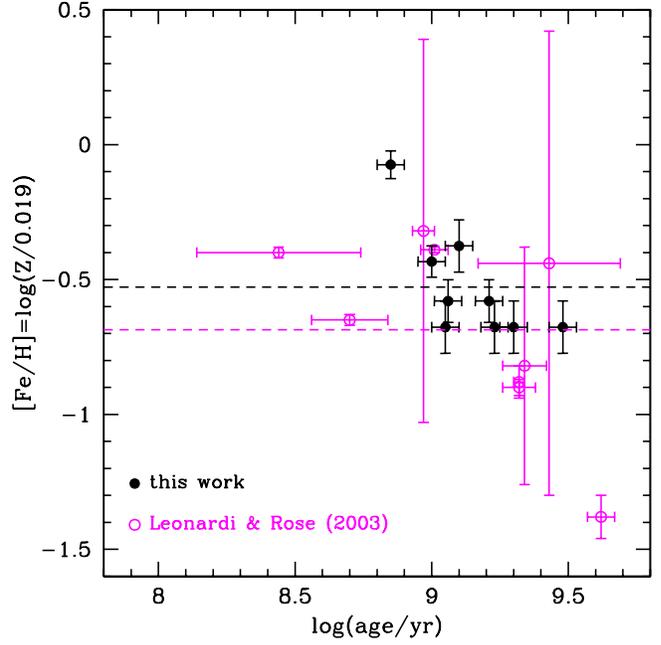}}
\caption{Age-metallicity relation from this work (black) in comparison
to that from Leonardi \& Rose (2003) (gray) for the same clusters. 
The dotted (dashed) line represents our (their) mean values.}
\label{KvsLR03_AMR}
\end{figure}

To better understand our contribution to the AMR, we selected the 
region covered by our cluster sample, comparing our results with 
the ones summarized by Mackey \& Gilmore (2003) and by 
Leonardi \& Rose (2003).
These comparisons are presented in Fig. \ref{KvsMG03_AMR} and 
Fig. \ref{KvsLR03_AMR}, respectively. 
Mackey \& Gilmore (2003) is a good compilation of results 
based mainly on ages determined by CMD analysis and metallicities 
from the OSSH spectroscopic study of red giants (with some
exceptions, as shown in Table \ref{lit_age_FeH}). 
Leonardi \& Rose (2003) is a study based on integrated spectra.
We are comparing here only results for  
clusters in common, which means 15 objects in the case of Mackey \& 
Gilmore (2003) and 9 objects in the case of Leonardi \& Rose (2003).

In general, our results have smaller uncertainties, especially 
when ages are concerned. 
Our results also have a lower spread in metallicities 
($\sigma=0.17$) when compared with those considered by 
Mackey \& Gilmore (2003) (from OSSH) ($\sigma=0.24$), 
and Leonardi \& Rose (2003) ($\sigma=0.34$).
Our mean metallicity in both comparisons is [Fe/H]\,$\sim-0.50$, 
lower than that from Mackey \& Gilmore (2003) ([Fe/H]\, 
$\sim-0.40$) and higher than the mean in Leonardi \& Rose (2003)
([Fe/H]\,$\sim-0.70$). 

\begin{figure}
\resizebox{\hsize}{!}{\includegraphics{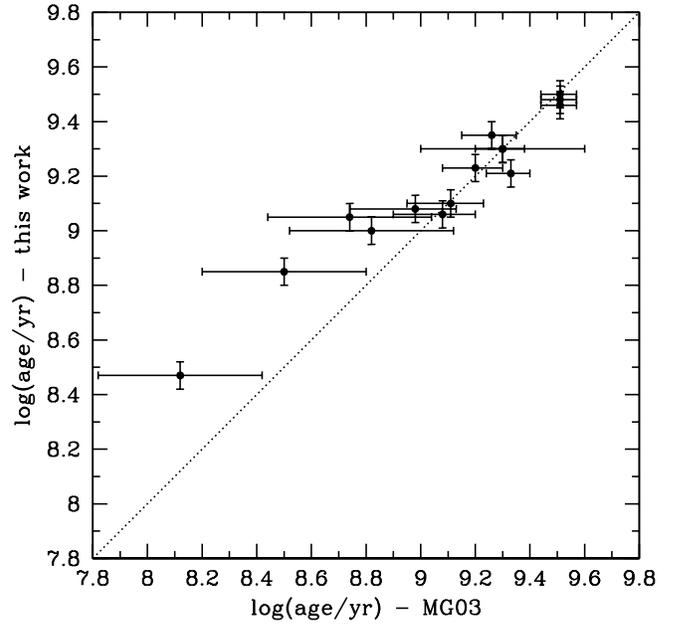}}
\caption{Our age results in comparison to the ones by Mackey 
\& Gilmore (2003) for the same clusters.
The identity relation is represented by the dotted line.}
\label{KvsMG03_age}
\end{figure}

\begin{figure}
\resizebox{\hsize}{!}{\includegraphics{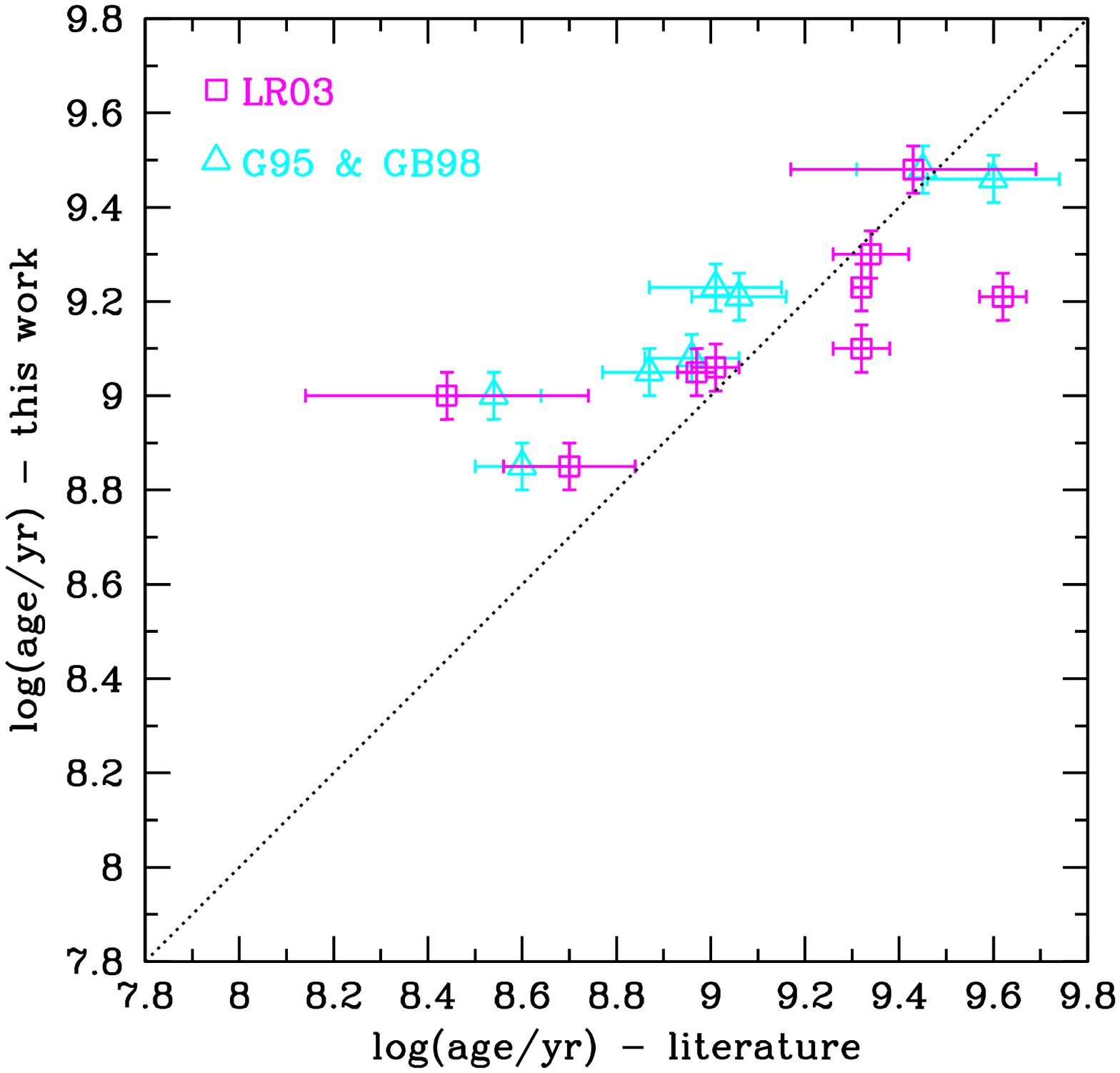}}
\caption{Our age results in comparison to the ones from 
Leonardi \& Rose (2003), Girardi et al. (1995) and Girardi \& Bertelli (1998).
The identity relation is represented by the dotted line.}
\label{Kvslit_age}
\end{figure}

In the following two figures we compare our age results with the ones 
presented by Mackey \& Gilmore (2003) (Fig. \ref{KvsMG03_age}) 
and by Leonardi \& Rose (2003), Girardi et al. (1995) and 
Girardi \& Bertelli (1998) (Fig. \ref{Kvslit_age}).
As discussed, the ages given by Mackey \& Gilmore (2003) 
come from CMD analysis, mainly using ground-based data 
(Elson \& Fall 1988; Geisler et al. 1997) but also using WFPC2/HST data 
(Rich et al. 2001) for the three oldest clusters. 
This comparison reveals a good agreement for clusters 
older than log($\tau$/yr)$\sim9.0$, especially for the ones
with more accurate photometry from WFPC2/HST. 
On the other hand, for clusters younger than this limit our 
results systematically predict higher ages, but are still 
consistent with the upper limit 
of the results compiled by Mackey \& Gilmore (2003). 
This systematic trend is observed when we compare our 
determinations with the ones from Girardi et al. (1995) and 
Girardi \& Bertelli (1998), whereas the Leonardi \& Rose (2003) 
results also present significant discrepancies for old clusters.
The trend towards higher age values for younger clusters
was recently reported by Beasley et al. (2002), 
where their results for the only cluster in common with us (NGC\,1856) 
seems to confirm this inference.

\begin{figure}
\resizebox{\hsize}{!}{\includegraphics{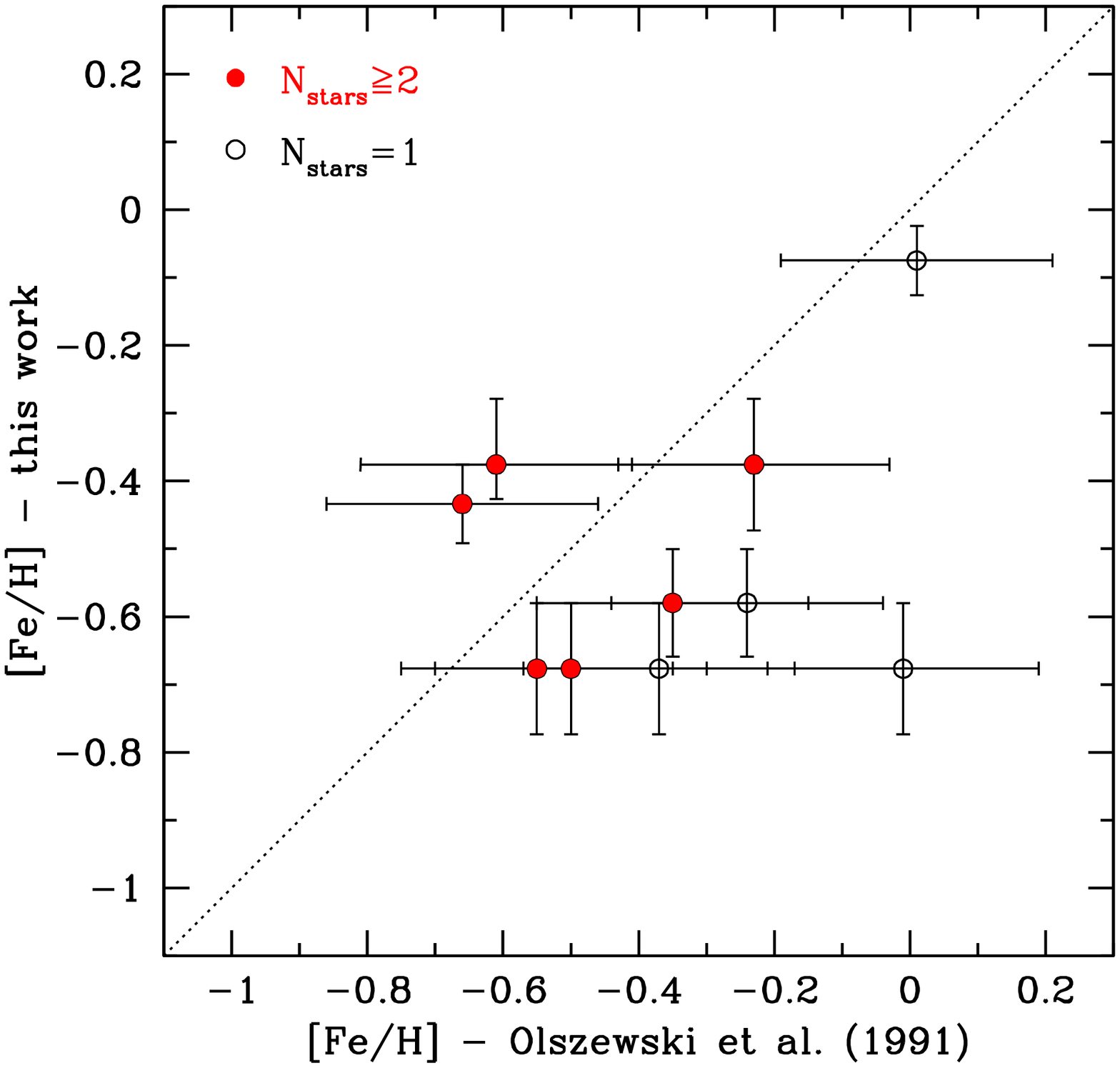}}
\caption{Our [Fe/H] values in comparison to the ones from OSSH.
The identity relation is represented by the dotted line.}
\label{KvsO91_FeH}
\end{figure}

Concerning metallicity, Fig. \ref{KvsO91_FeH} contrasts our results
with the ones from OSSH (also selected by Mackey \& Gilmore 2003).
A preliminary comparison indicates some discrepancies close to 
[Fe/H] $\sim-0.70$ (in our results), in the sense that OSSH 
determined higher values than us. 
However, these most discrepant points (NGC\,1651, NGC\,2173 and NGC\,2213)
come from determinations where only one star per 
cluster was used, therefore being subject to high uncertainties.
Geisler recently showed at a FONDAP/ESO Conference 
(Globular Cluster - Guide to Galaxies) (Geisler 2006; Grocholski et al. 2006) 
new results for these clusters in agreement with our [Fe/H] values.
We also notice that the recalibrated OSSH values for the 
correcting transformation proposed by Cole et al. (2005)
tend to slightly reduce the differences between metallicities 
(the mean difference between us and OSSH,  
$<\rm{this~work-OSSH}>_{\rm{[Fe/H]}}$,  drops from -0.16 to -0.10). 
Considering only the OSSH determinations based on more than one 
star per cluster,
the mean metallicity becomes very similar in both samples
($\sim 0.50$, with an insignificant mean difference of 
$<\rm{this~work-OSSH}>_{\rm{[Fe/H]}}\sim$ -0.02) and insensitive to the 
correction proposed by Cole et al. (2005).

\begin{figure}
\resizebox{\hsize}{!}{\includegraphics{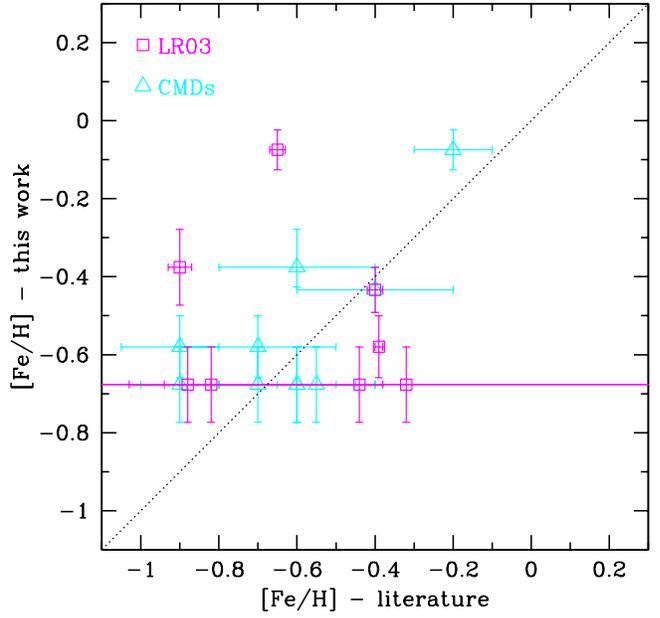}}
\caption{Our [Fe/H] values in comparison to the ones from CMDs
and from Leonardi \& Rose (2003).
The identity relation is represented by the dotted line.}
\label{Kvslit_FeH}
\end{figure}

Comparisons of our metallicity results with the ones obtained 
from CMD analysis and by Leonardi \& Rose (2003) are shown in 
Fig. \ref{Kvslit_FeH}.
While the results based on CMDs reveal a satisfactory agreement, 
the Leonardi \& Rose (2003) ones present some discrepant 
points, in the sense that they determined significantly lower 
metallicities.  However, the quoted uncertainties by these latter 
authors may be an underestimate, specially for a 
study based on integrated spectra. 

\begin{figure}
\resizebox{\hsize}{!}{\includegraphics{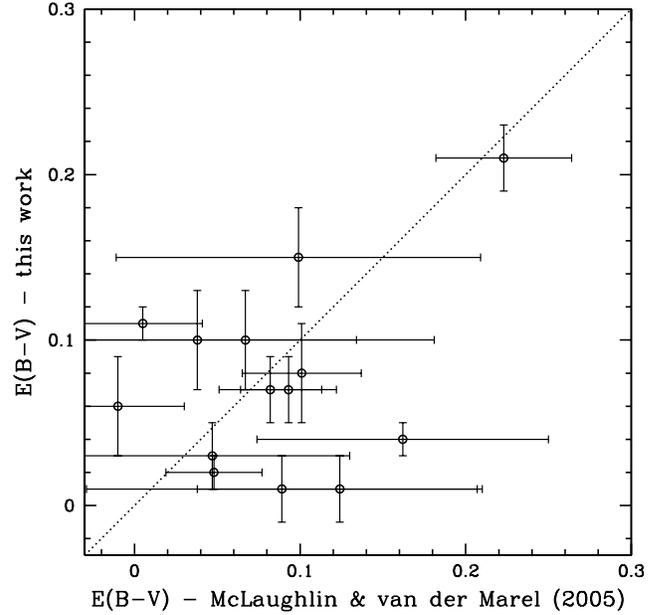}}
\caption{Our reddening results in comparison to the ones from 
McLaughlin \& van der Marel (2005).
The identity relation is represented by the dotted line.}
\label{KvsMv05_EBV}
\end{figure}

We compare our reddening results with the ones obtained 
by McLaughlin \& van der Marel (2005) for the 14 clusters we have 
in common.
These comparisons, shown in Fig. \ref{KvsMv05_EBV}, reveal larger
uncertainties in the McLaughlin \& van der Marel (2005) estimates, 
jeopardizing a more systematic comparison between them.
However, our most discrepant reddening value 
(that of NGC\,1856, which is significantly higher than the others)
is in accordance with McLaughlin \& van der Marel (2005). 

\begin{figure}
\resizebox{\hsize}{!}{\includegraphics{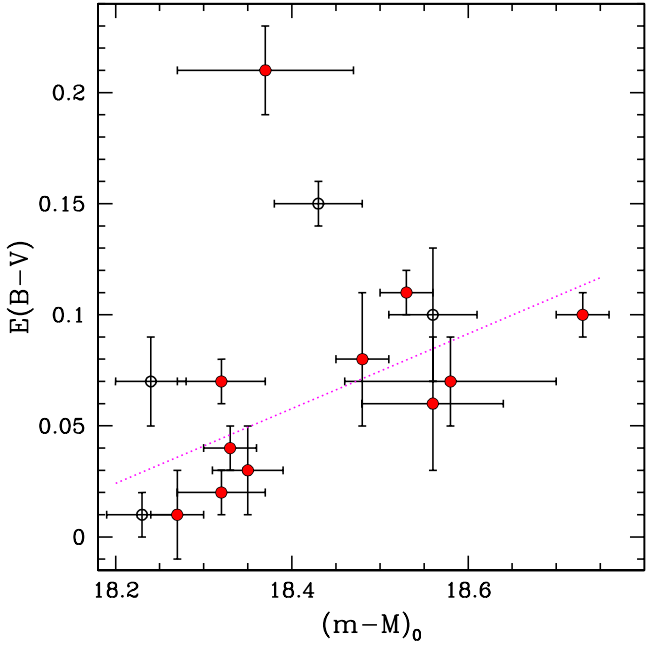}}
\caption{
Relation between our inferred reddening and distance modulus.
Solid (open) circles are the best models confined to 
$n \le 3$ ($n > 3$), and therefore represent the most (least) 
reliable ones.  
The dotted line is the linear fit for the most reliable solutions
excluding the highest reddening value.
}
\label{ebv_vs_modist}
\end{figure}

Since a relation between reddening and distance modulus is expected,
we plot our results for these two parameters in 
Fig. \ref{ebv_vs_modist}.
Except for some outlying points and less reliable estimates 
(open circles) a clear and consistent trend is revealed: more 
distant clusters tend to be more reddened and vice-versa, as 
represented by the linear fit shown in this figure. 
Using this derived relation, a typical E(B-V) value for the
LMC is found ($\sim 0.07$) for the canonical LMC distance 
($(m-M)_0$=18.50). This reddening value has an amplitude of 
$\sim 0.05$, possibly reflecting the variations in the LMC 
optical depth.

\begin{figure}
\resizebox{\hsize}{!}{\includegraphics{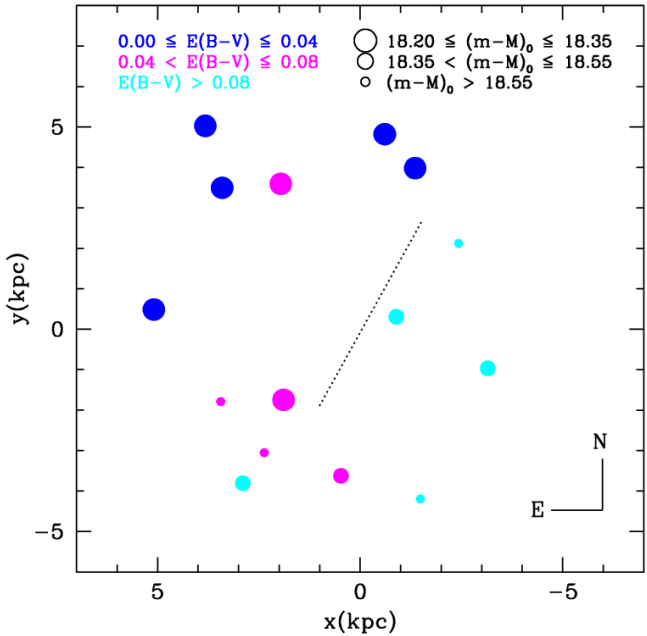}}
\caption{
Distribution of LMC clusters on the sky in physical units (as in Fig.
\ref{LMCsky}, using the relation $1\degr=0.873$ kpc corresponding
to the $(m-M)_{0}=18.50$), with symbols coded according with 
their reddening (gray scale) and distance modulus (size).
The dotted line indicates the line of nodes for the disk, 
as determined by Nikolaev et al. (2004).
The distances are relative to the optical centre of the LMC bar
(Bica et al. 1996)
}
\label{LMCsky_modist_ebv}
\end{figure}

\begin{figure}
\resizebox{\hsize}{!}{\includegraphics{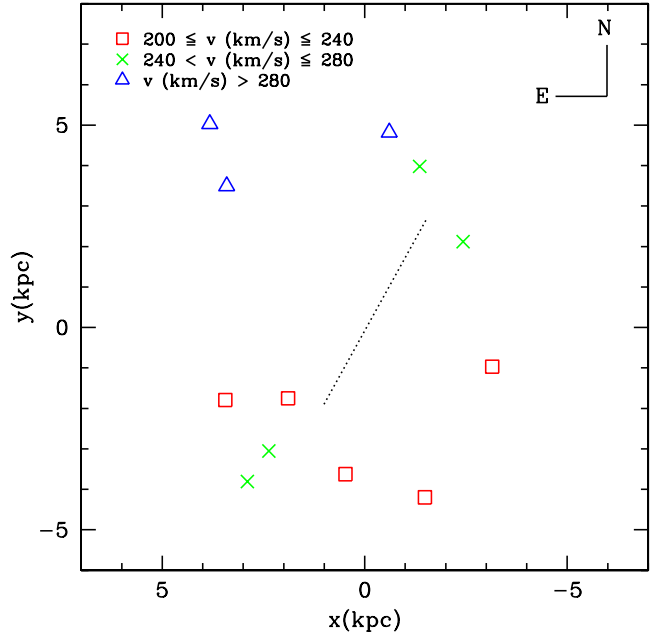}}
\caption{
Same as in Fig. \ref{LMCsky_modist_ebv}, but 
with symbols coded according with their velocity, 
as determined by OSSH.
}
\label{LMCsky_veloc}
\end{figure}

The on-sky distribution of the clusters  
(Figs. \ref{LMCsky_modist_ebv}-\ref{LMCsky_veloc}) reveals 
that the lowest reddening, closest and highest velocity 
(as determined by OSSH) ones are 
preferentially located in the NE region.
On the other hand, the SW region seems to host clusters with 
the opposite characteristics.

\begin{figure}
\resizebox{\hsize}{!}{\includegraphics{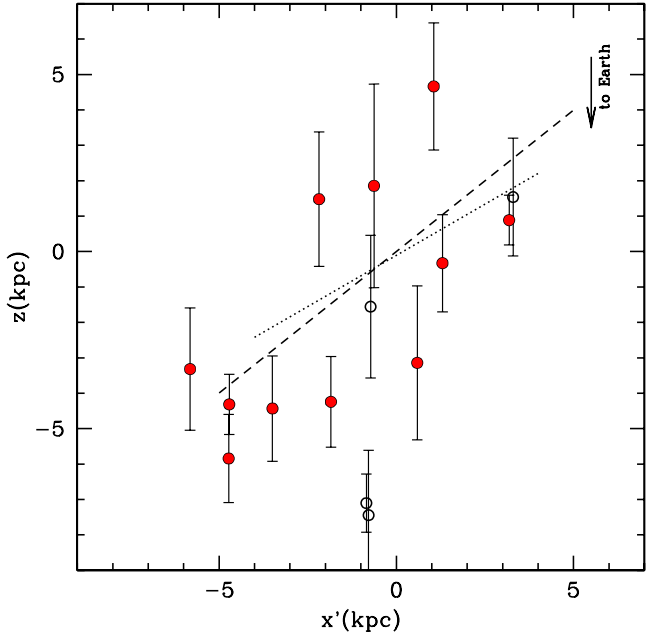}}
\caption{
Distribution of LMC clusters as seen by an imaginary observer 
whose $z$-axis is along the line of nodes. 
The dotted line indicates the projected LMC disk plane 
as determined by Nikolaev et al. (2004). 
The linear solution for the clusters with the 
most reliable results is shown by the dashed line.
The direction to Earth is shown in the up right corner.
}
\label{LMCsky_z_vs_nx}
\end{figure}

This result for the spatial distribution of clusters 
agrees well with the geometry known for the LMC disk 
(Westerlund 1997) and recently revisited by Nikolaev et al. (2004) 
using Cepheids, who determined a position angle ($\theta$) 
of $151.0\degr \pm 2.4\degr$ for the line of nodes and 
$30.7\degr \pm 1.1 \degr$ 
for the disk inclination ($i$), with the northeast quadrant 
being the closest. 
Fig. \ref{LMCsky_z_vs_nx} illustrates this accordance
showing the cluster positions on a plane for an imaginary observer 
with the line of sight aligned with the line of nodes for the disk. 
In other words, the $z$ axis for this plane is at the canonical 
distance to the LMC centre 
($(m-M)_{0}$=18.50), while the $x'$ axis is the 
perpendicular distance to the line of nodes.
Although the cluster distribution is scattered, the clusters 
with the most reliable determinations have an inclination 
of $\sim 39\degr$ ($\pm 7\degr$) and thus roughly follow the disk, 
with the closest clusters at the NE quadrant 
(negative $x'$ values).
The mean distance modulus calculated for the whole cluster sample 
is 18.42 (with a dispersion of 0.16), being slightly lower than the 
typical assumed LMC distance, likely reflecting a larger number of
systems located in the NE quadrant than in the SW one. 
If we consider only the most reliable solutions, the mean and dispersion 
for the distance modulus remain almost the same, being 18.44 and 0.14,
respectively.

\begin{figure}
\resizebox{\hsize}{!}{\includegraphics{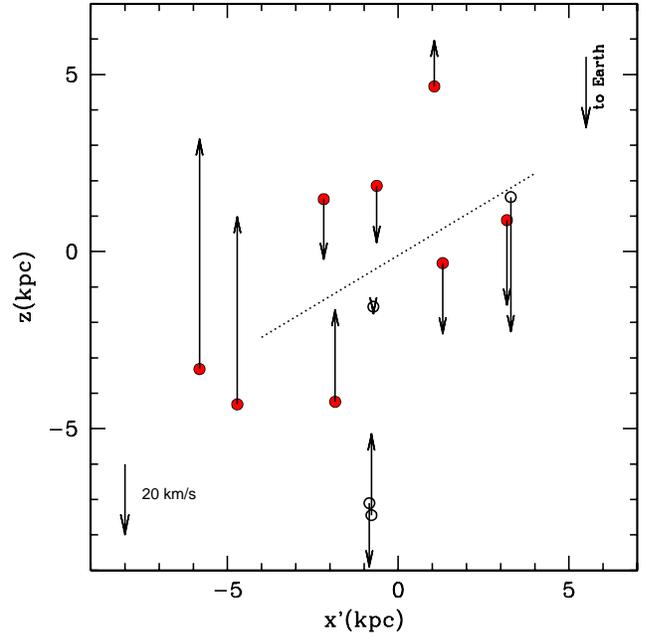}}
\caption{
Same as Fig. \ref{LMCsky_z_vs_nx}, but showing the radial velocity 
of each cluster, as determined by OSSH, relative to the mean radial 
velocity of stars belonging to the LMC bar as determined by 
Cole et al. (2005). 
A velocity scale is shown in the bottom left corner.  
}
\label{LMCsky_z_vs_nx_veloc}
\end{figure}

The radial velocities determined by OSSH relative to the mean 
radial velocity determined by Cole et al. (2005) (256 km/s) are 
represented in this plane 
in Fig. \ref{LMCsky_z_vs_nx_veloc}.
This figure is consistent with
a clockwise rotation curve, with the extreme NE ($-x'$) or SW ($+x'$)
clusters with the highest absolute values for the radial velocities. 
This rotation curve, combined with the derived inclination for the 
cluster distribution, suggests a (slightly) kinematically delayed
structure compared to the bulk of stars in the LMC disk.

\begin{figure}
\resizebox{\hsize}{!}{\includegraphics{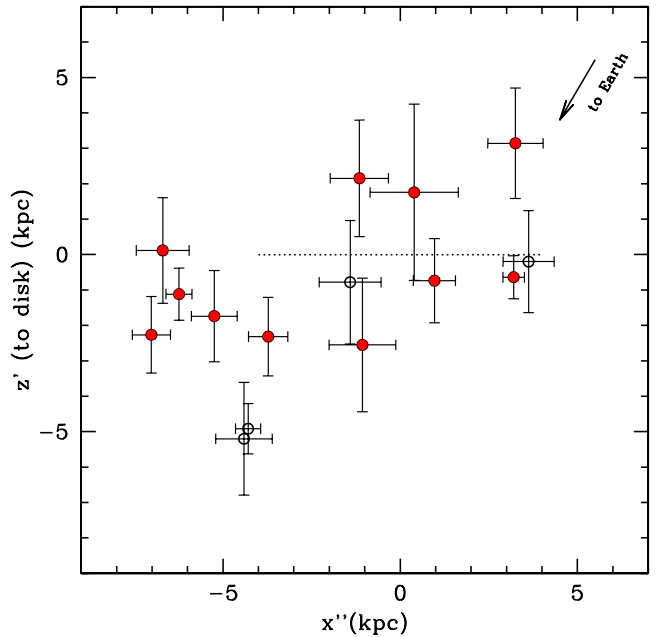}}
\caption{
Spatial distribution for the LMC clusters in a plane aligned to
the LMC disk (indicated by the dotted line). 
The direction towards us is shown in the top-right corner. 
}
\label{LMCsky_nz_vs_nnx}
\end{figure}

A plane with axes aligned with the LMC disk 
(the $z$-$x'$ plane rotated clockwise by $i$) is presented 
in Fig. \ref{LMCsky_nz_vs_nnx}.
Almost all clusters with the most accurate modelling
are confined to a distance perpendicular to the disk plane lower 
than 3 kpc. 

\section{Summary and conclusions}
\label{end}

We analyzed HST/WFPC2 CMDs from 15 LMC populous clusters 
to determine the following physical parameters for each of them:
age, metallicity, distance modulus and reddening.
For each cluster, the observed MS fiducial line and RC position
 were simultaneously and statistically compared 
with the ones obtained from synthetic CMDs. 
The CMD models explored a regular grid in the parameter space 
consistent with previous determinations found in the literature. 
Control experiments were used to test our approach. 
Therefore, our determinations, based on photometrically
homogeneous data, are self-consistent and done by an objective and 
robust method. 

In general, the best models show a satisfactory fit to the data,
reproducing well the MS fiducial line and RC features. 
Also, these models constrain well the physical parameters of 
each cluster, with typical uncertainties of 0.05 in log($\tau$/yr), 
0.10 dex in [Fe/H], 0.05 in $(m-M)_{0}$ and 0.02 in $E(B-V)$.
These results are summarized in Table \ref{parameters}.

The AMR derived from our results has a lower spread in metallicity 
than the one compiled by Mackey \& Gilmore (2003) or the one 
recently determined by Leonardi \& Rose (2003) using integrated 
colors.
We also recovered a mean [Fe/H] of $\sim$ -0.50, roughly 
$\sim 0.10$ more metal-poor than the Mackey \& Gilmore (2003) data 
and $\sim 0.20$ more metal-rich than Leonardi \& Rose (2003).
 
The metallicity values determined by us are in accordance with the 
ones from OSSH where more than one star per cluster was used to 
measure the [Fe/H] cluster value. 
The uncertainties in our [Fe/H] estimates are 
comparable with the ones obtained by OSSH, with the 
advantage that we recovered it based on a statistical
method.  
Comparisons with previous metallicities determined using
CMDs also reveal a good agreement, but those based on integrated 
light show some discrepant points.
 
In terms of age, the earlier determinations based on CMDs 
are in good agreement with ours for clusters older than 
log($\tau$/yr)$\sim9.0$. Below this limit, earlier 
results systematically recovered younger 
ages.
The results from Leonardi \& Rose (2003) based on integrated 
light also show discrepancies for old clusters, in the 
sense that they obtained an older age than us. 

In general, reddening and distance modulus 
have canonical values adopted in the previous studies, 
often not being consistently determined 
individually for each cluster. 
An exception is the E(B-V) from McLaughlin \& van der Marel (2005), 
a study based on integrated colours, but 
their values have large uncertainties and do not allow a systematic 
comparison to our results. 
For NGC\,1856, however, the cluster with the highest 
E(B-V)($\sim 0.21$) value in our work, those authors also found 
their highest reddening value, in good agreement with us.

A consistent and expected relation involving reddening and distance 
modulus was found, in the sense that clusters with lower extinction 
tend to be in the foreground.
The three-dimensional distribution of the clusters with the most 
reliable results seems to be roughly aligned with the LMC disk geometry,
with a small difference of $\sim 8 \degr$ in the inclination, 
suggesting a (slightly) kinematically delayed structure for the 
system composed of the IACs in relation to the bulk of LMC disk stars.
Although these clusters are restricted to a distance perpendicular 
to the disk lower than 3kpc, they seem to be more scattered than 
the numerical predictions for the formation and evolution of 
intermediate-age LMC clusters done by Bekki \& Chiba (2005).
Therefore, the results of the three-dimensional distribution of the 
IACs in the LMC may be interpreted as an indication that these clusters 
were not formed in the LMC disk. Alternatively, they 
may have formed in the disk but been scattered away from it by
interactions as they moved through the LMC potential.

We underline that the set of age and metallicities
homogeneously derived here can be applied to calibrated light studies of
distant galaxies. 
Since our results are based on the Padova models, it would also be very
interesting to work with the same data to allow the intercomparison 
of predictions based on other stellar evolutionary models, 
like the Y$^2$ (Yi et al. 2003), the Pisa (Castellani et al. 2003) 
and the Teramo (Pietrinferni et al. 2004) ones.
Concerning these last models a quantitative result using 
synthetic CMDs to derive ages and reddenings of a small sample 
of LMC star clusters has been obtained by Raimondo et al. (2005) 
for a different purpose.

\begin{acknowledgements}
We thank Leo Girardi for kindly provide a thinner 
isochrone grid in metallicity.
LOK thanks Beatriz Barbuy for useful discussions and 
acknowledges FAPESP postdoctoral fellowship 05/01351-5. 

\end{acknowledgements}

\begin{appendix}

\section{Control Experiments}
\label{appendix}

This appendix shows some results of control 
experiments, used to test our approach to determine the physical 
parameters of a cluster.
For each control experiment, a synthetic CMD with known (input) parameters
is assumed as an ``observed CMD'' and compared with a regular model grid.
Table \ref{experiments} presents the results (output) of a sample of such 
experiments, numbered from 1 to 4, for a fixed n=2, and originally
designed to quantify the formal uncertainties in the results for 
NGC\,1831, NGC\,1868, NGC\,2173 and NGC\,2121. 
As expected, the input parameters are recovered in the output for all
experiments, attesting the applicability of our statistical tools.  
This table reveals that single criteria (MS or RC) have 
output parameters with higher uncertainties than the ones recovered
when the both criteria are combined.
These uncertainties are directly related to the number of models 
identified as best models ($N_{best}$), as attested by the last column in 
this table.

\begin{table*}
\caption{Control Experiments}
\label{experiments}
\renewcommand{\tabcolsep}{1.5mm}
\centering
\begin{tabular}{l c c c c c c c c c c}
\hline\hline
ID & \multicolumn{4}{c}{input} & criterion & \multicolumn{5}{c}{output} \\
~~~ & log($\tau$/yr) & $Z$ & $(m-M)_{0}$ & $E(B-V)$  & ~~~ & log($\tau$/yr) & $Z$ & $(m-M)_{0}$ & $E(B-V)$ & $N_{best}$ \\
\hline
1 & 8.85 & 0.016 & 18.25 & 0.01 & MS & $8.81\pm0.03$ & $0.017\pm0.002$ & 
$18.30\pm0.09$ & $0.03\pm0.02$ & 83 \\
 &  &  &  &  & RC & $8.83\pm0.04$ & $0.016\pm0.002$ & 
$18.28\pm0.10$ & $0.02\pm0.02$ & 39 \\
 &  &  &  &  & MS\&RC & $8.84\pm0.02$ & $0.016\pm0.002$ & 
$18.26\pm0.07$ & $0.02\pm0.02$ & 21 \\
2 & 9.05 & 0.004 & 18.35 & 0.04 & MS & $9.03\pm0.05$ & $0.005\pm0.001$ & 
$18.39\pm0.12$ & $0.04\pm0.03$ & 118 \\
 &  &  &  &  & RC & $9.02\pm0.06$ & $0.004\pm0.000$ & 
$18.42\pm0.09$ & $0.04\pm0.01$ & 25 \\
 &  &  &  &  & MS\&RC & $9.07\pm0.03$ & $0.004\pm0.000$ & 
$18.34\pm0.03$ & $0.04\pm0.01$ & 8 \\
3 & 9.20 & 0.004 & 18.60 & 0.07 & MS & $9.20\pm0.06$ & $0.005\pm0.001$ & 
$18.56\pm0.12$ & $0.06\pm0.03$ & 244 \\
 &  &  &  &  & RC & $9.18\pm0.05$ & $0.005\pm0.001$ & 
$18.59\pm0.14$ & $0.05\pm0.03$ & 24 \\
 &  &  &  &  & MS\&RC & $9.20\pm0.04$ & $0.004\pm0.000$ & 
$18.62\pm0.10$ & $0.06\pm0.02$ & 12 \\
4 & 9.45 & 0.008 & 18.25 & 0.07 & MS & $9.43\pm0.05$ & $0.008\pm0.002$ & 
$18.29\pm0.10$ & $0.07\pm0.03$ & 82 \\
 &  &  &  &  & RC & $9.45\pm0.10$ & $0.008\pm0.002$ & 
$18.25\pm0.06$ & $0.08\pm0.03$ & 48 \\
 &  &  &  &  & MS\&RC & $9.44\pm0.04$ & $0.008\pm0.001$ & 
$18.24\pm0.05$ & $0.08\pm0.02$ & 11 \\
\hline
\end{tabular}
\end{table*} 

\end{appendix}

\end{document}